\documentclass[aps,pra,twocolumn,superscriptaddress,showpacs]{revtex4-1}
\usepackage{graphicx}
\usepackage{dcolumn}
\usepackage{bm}
\begin{document}

\title{Nonlinear Bose-Einstein-condensate Dynamics Induced by a Harmonic\\
Modulation of the \textit{\textbf{s}}-wave Scattering Length}

\author{Ivana Vidanovi\' c}
\email[]{ivana.vidanovic@ipb.ac.rs}
\homepage[]{http://www.scl.rs/}
\affiliation{Scientific Computing Laboratory,
Institute of Physics Belgrade,
University of Belgrade,
Pregrevica 118, 11080 Belgrade, Serbia}

\author{Antun Bala\v z}
\affiliation{Scientific Computing Laboratory,
Institute of Physics Belgrade,
University of Belgrade,
Pregrevica 118, 11080 Belgrade, Serbia}

\author{Hamid Al-Jibbouri}
\affiliation{Institut f\"{u}r Theoretische Physik, Freie Universit\"{a}t Berlin,
Arnimallee 14, 14195 Berlin, Germany}

\author{Axel Pelster}
\affiliation{Fachbereich Physik, Universit\"{a}t Duisburg-Essen, Lotharstrasse 1, 47048 Duisburg, Germany}
\affiliation{Fachbereich Physik, Universit\"{a}t Bielefeld, Universit\"{a}tsstrasse 25, 33501 Bielefeld, Germany}

\date{\today}

\begin{abstract}
In a recent experiment, a Bose-Einstein condensate of ${}^7$Li
has been
excited by a harmonic modulation of the atomic \textit{s}-wave scattering length
via Feshbach resonance. By combining an analytical perturbative approach with
extensive numerical simulations we analyze the emerging nonlinear dynamics of the
system on the mean-field Gross-Pitaevskii level at zero temperature. Resulting excitation spectra are presented and  prominent nonlinear
features are found: mode coupling, higher harmonics generation and
significant shifts in the frequencies of collective modes. 
We indicate how nonlinear dynamical properties 
could be made clearly observable in future experiments and compared to our results.
\end{abstract}

\pacs{03.75.Kk, 03.75.Nt, 67.85.De}

\maketitle
\section{Introduction}

Soon after the first experimental realization of an atomic Bose-Einstein condensate (BEC), the
understanding of the physical properties of the system has been tested by comparing experimental results for the frequencies of collective  excitation modes \cite{debbie,ketterle} with corresponding theoretical results \cite{linearresponsegp, stringari, variational, variational2, zaremba}. The same paradigm remains of great relevance for  present and future studies that focus on a BEC in optical lattices \cite{stronglycorrelated}, in  disordered potentials \cite{disorder}, for low-dimensional cold bosons \cite{stronglycorrelated,lowdim} or ultracold fermions \cite{coldfermi, lima-fermions}. The essential merit of testing theoretical predictions using collective modes stems from the possibility to measure frequencies of collective modes with a high accuracy of the order of $1\%$ \cite {acc1, acc2}. For instance, a promising experimental proposal for observing a quantum anomaly based on a very precise measurement of the collective BEC mode, has been recently suggested in Ref.~\cite{olshanii}.

Usually, collective modes are induced by a modulation of the external trapping potential \cite{debbie, ketterle, timedeptrap1, timedeptrap2, timedeptrap3, frequencyshift}. In the recent paper \cite{parres},  however, an alternative way of a condensate excitation has been experimentally realized. A broad Feshbach resonance of ${}^7$Li \cite{Feshbach} allows to modulate the atomic \textit{s}-wave
scattering length by modulating an external magnetic field. Based on this property, a harmonic modulation of the \textit{s}-wave
scattering length has been achieved
\begin{equation}
a(t)\simeq a_{\mathrm{av}}+\delta_a \cos\Omega t,
\label{eq:tdsl}
\end{equation}
with $a_{\mathrm{av}}\approx3 a_0>\delta_a\approx 2 a_0$, where $a_0$ is the Bohr radius, yielding a time-dependent interaction among atoms. For the specific set of experimental parameters, a quadrupole oscillation mode has been excited in this way and resonances located at the quadrupole
frequency and its second harmonic have been observed.
There are several advantages of such experimental scheme: for instance, in future experiments with multispecies BEC, a single component could be individually excited in this way while the excitation of other components would occur only indirectly.

Motivated by the latter experimental study, in this paper we analyze the condensate dynamics induced by the harmonic modulation of the \textit{s}-wave scattering length in more detail. On the mean-field level at zero temperature, the BEC dynamics is captured by the nonlinear Gross-Pitaevskii (GP) equation for the condensate wavefunction and the time-dependent interaction leads to a time-dependent nonlinearity. Depending on the closeness of the external modulation frequency $\Omega$ to one of condensate's eigenmodes, a qualitatively different dynamical behavior emerges. In the non-resonant case, we have small-amplitude oscillations of the condensate size around the equilibrium widths, and we are in the regime of linear response.  However, as $\Omega$ approaches an eigenmode, we expect a resonant behavior which is characterized by large amplitude oscillations.
In this case it is clear that a linear response analysis does not provide a qualitatively good description of the system dynamics. 

First theoretical studies of collective modes of BEC \cite{linearresponsegp, stringari, variational} have explored dynamical properties in the linear regime of small amplitude oscillations. Certain nonlinear aspects of a condensate dynamics induced by a trap modulation are given in Refs.~\cite{frequencyshift, timedeptrap2, timedeptrap3}, whereas two-component BECs are dealt with in Refs.~\cite{twocomponentnonlinearities1, twocomponentnonlinearities2}. We emphasize that very small oscillation amplitudes, which occur in the linear regime, are experimentally hard to observe. On the other hand, very large amplitude oscillations lead to a fragmentation of the condensate \cite{parres, yukalov}. Thus, the case in between is of main experimental interest and represents the main objective of our study.

To this end, we use an approach that is complementary to the recent theoretical considerations \cite{parresnumericsAdhikari1, malomed, yukalov2, Abdullaev1, Abdullaev2,staliunas} of a BEC with harmonically modulated interaction. In Ref.~\cite{Abdullaev1} the real-time dynamics of a spherically symmetric BEC was numerically studied and analytically explained using the resonant Bogoliubov-Mitropolsky method \cite{book3}. On the other hand, in our approach in order to discern induced dynamical features, we look directly at the excitation spectrum obtained as a Fourier transform of the time-dependent condensate width.  From this type of numerical analysis we find characteristic nonlinear properties: higher harmonic generation, nonlinear mode coupling, and significant shifts in the frequencies of collective modes with respect to linear response results. In addition, we work out an analytic perturbative theory with respect to the modulation amplitude capable of capturing many of the mentioned nonlinear effects obtained numerically. 

Nonlinearity-induced frequency shifts were considered previously in Ref.~\cite{frequencyshift}
for the case of bosonic collective modes excited by modulation of the trapping potential,
and also in Ref.~\cite{chineseprb} for the case of a superfluid Fermi gas in the BCS-BEC crossover.
Our analytical approach is based on the Poincar\'e-Lindstedt method \cite{poinlind, book1, book2, book3}, in the same spirit as presented in Refs.~\cite{frequencyshift, chineseprb, poinlind}. However, the harmonic modulation of the interaction strength introduces additional features that require a separate treatment.

In Ref.~\cite{staliunas} it was predicted that a harmonic modulation  of the scattering length  leads to the creation of Faraday patterns, i.e. density waves, in BEC. Up to now, Faraday patterns have been  experimentally induced by harmonic modulation of the transverse confinement strength \cite{Faradaypatternsexp}, which is studied analytically and numerically in Ref.~\cite{Faradaypatternstheo}. In this paper we focus only on the nonlinear properties of low-lying collective modes and do not consider possible excitations of Faraday patterns.

The paper is organized in the following way: in Sec.~II we provide the necessary theoretical background - the mean-field GP equation and the time-dependent variational approximation of its solution. In Sec.~III we consider a spherically symmetric BEC with harmonically modulated scattering length. We analyze the condensate dynamics in detail and identify characteristic nonlinear features. In Sec.~IV we focus on the experimentally most interesting axially symmetric BEC. Finally, in Sec.~V we summarize the main points of our study and discuss the relevance of our results for future experimental setups.

\section{Method}

To study the dynamics of an atomic BEC at zero temperature, we
use the mean-field description given by the time-dependent GP equation \cite{bookgp}
\begin{equation}
i\hbar\frac{\partial \psi(\vec{r},t)}{\partial t}=\left[-\frac{\hbar^2}{2
m}\Delta + V(\vec{r}) + g |\psi(\vec{r},t))|^2 \right]\psi(\vec{r},t),
\label{eq:gp}
\end{equation}
where $\psi(\vec{r},t)$ is the condensate wave-function normalized to unity. On the right-hand side we have a kinetic energy term, an external axially symmetric
harmonic trap potential $
V(\vec{r})=\frac{1}{2}m\omega_{\rho}^2(\rho^2+\lambda^2 z^2)$ with the trap aspect ratio  $\lambda$, and a nonlinear term arising from the mean-field interaction between the atoms. 
The interaction strength $g=\frac{4 \pi
\hbar^2 N a}{m}$  is proportional to the total number of atoms in the condensate
$N$ and to the  \textit{s}-wave scattering length $a$. A specific feature of the recent experiment \cite{parres} is the harmonic modulation of
the \textit{s}-wave scattering length described by Eq.~(\ref{eq:tdsl}),
yielding a time-dependent interaction strength $g=g(t)$.  

In order to obtain analytical insight into the condensate dynamics induced in this way,
we use the Gaussian approximation introduced in Refs.~\cite{variational,variational2}. To this end we assume that the condensate wave function with contact interaction has the same Gaussian form as in the non-interacting case, just with renormalized parameters. Thus, we use a time-dependent variational method based on a Gaussian ansatz, which reads for an axially symmetric trap:
\begin{eqnarray}
\psi^G(\rho,z,t)&=&{\mathcal N}(t) \exp\left[-\frac{1}{2}\frac{\rho^2}{u_{\rho}(t)^2}+i
\rho^2 \phi_{\rho}(t)\right]\nonumber\\ 
&\times&\exp\left[-\frac{1}{2}\frac{z^2}{u_z(t)^2}+i z^2 \phi_ z (t)\right],
\label{eq:gauss}
\end{eqnarray}
where ${\mathcal N}(t)=\pi^{-\frac{3}{4}} u_{\rho}(t)^{-1} u_z(t)^{-\frac{1}{2}}$ is a time-dependent normalization, while $u_{\rho}(t)$, $u_{z}(t)$,
$\phi_z(t)$ and $\phi_{\rho}(t)$ are variational parameters. 
They have straightforward interpretation: $u_\rho(t)$ and $u_z(t)$ correspond to the radial and to the axial condensate width, while $\phi_\rho(t)$ an $\phi_z(t)$ represent the corresponding phases. Therefore, the above ansatz describes dynamics of the condensate in terms of the time-dependent condensate widths and phases, while no center of mass motion is considered here. This is due to the fact that modulation of the interaction strength does not induce this type of motion.

Following the variational approach introduced earlier in Ref.~\cite{variational}, we insert ansatz (\ref{eq:gauss}) to the action yielding the GP equation, and extremize it with respect to variational parameters. In this way we obtain a system of ordinary differential equations that govern the condensate dynamics.
Although here we consider a time-dependent interaction, this does not change the main steps in the derivation of variational equations. By extremizing the action, we first obtain a coupled system of differential equations of the first order for all variational parameters. The equations for the phases $\phi_\rho$ and $\phi_z$ can be solved explicitly in terms of widths $u_\rho$ and $u_z$, yielding
\begin{equation}
 \phi_{\rho}=m\dot{u}_{\rho}/(2 \hbar u_{\rho}),\quad\phi_z=m\dot{u}_z/(2 \hbar u_z)\, .
\end{equation}
After inserting the above expressions into equations for the widths, we finally obtain a system of two differential equations of second order for $u_\rho$ and $u_z$, which we will refer to as a Gaussian approximation:

\begin{eqnarray}
 \ddot{u}_{\rho}(t)+u_{\rho}(t)-\frac{1}{u_{\rho}(t)^3}-\frac{p(t)}{u_{\rho}(t)^3
u_z(t)}&=&0,\label{eq:var1}\\
\ddot{u}_z(t)+\lambda^2 u_z(t)-\frac{1}{u_z(t)^3}-\frac{p(t)}{u_{\rho}(t)^2
u_z(t)^2}&=&0.
\label{eq:var2}
\end{eqnarray}
In the previous set of equations and in all equations that follow, we express all lengths in the units of the characteristic harmonic oscillator length $l=\sqrt{\hbar/m\omega_{\rho}}$ and time in units of $\omega_{\rho}^{-1}$. 
The dimensionless interaction parameter $p(t)$ is given by $p(t)= \sqrt{2/\pi} N a(t)/l $.

Taking into account Eq.~(\ref{eq:tdsl}), we have:
\begin{equation}
 p(t)=p+q \cos \Omega t,
\label{eq:modint}
\end{equation}
where $p=\sqrt{2/\pi} N a_{\mathrm {av}}/l$ denotes the average interaction strength, $q=\sqrt{2/\pi} N \delta_a/l$ is a modulation amplitude, and $\Omega$ represents the modulation or driving frequency.

\begin{figure}[!hbt]
\begin{center}
\includegraphics[width=8cm]{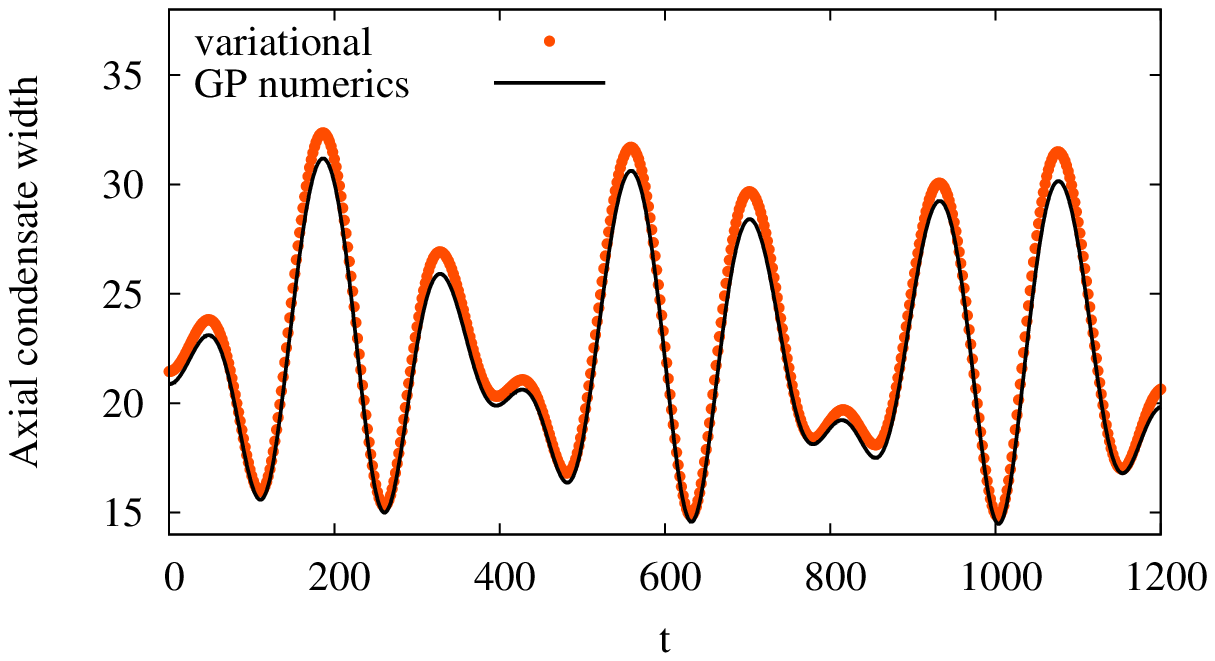}
\includegraphics[width=8cm]{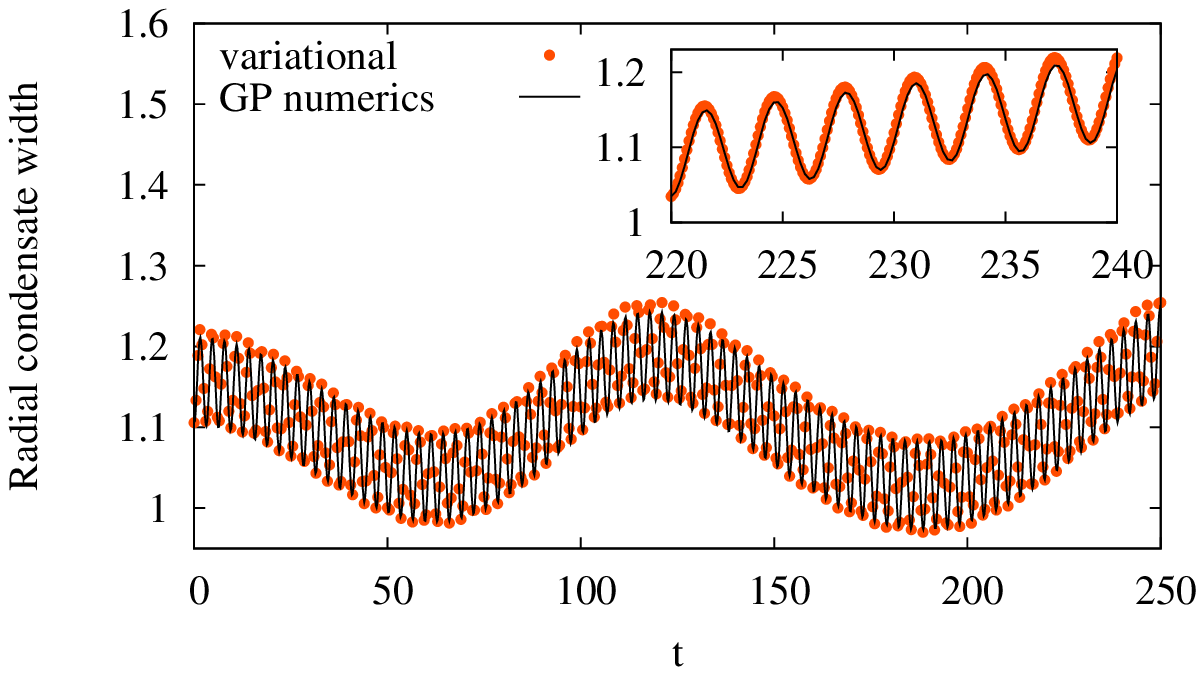}
\caption{(Color online) Time-dependent axial and radial condensate widths calculated as root mean square averages. Comparison of the numerical solution of time-dependent GP equation with a solution obtained by using the Gaussian approximation for the actual experimental parameters in Eq.~(\ref{eq:exppar}) and $\Omega=0.05$.}
\label{fig:gp}
\end{center}
\end{figure}

To estimate the accuracy of the Gaussian approximation for describing the dynamics induced by the harmonic modulation of the interaction strength, we compare it with an exact numerical solution of GP equation. The experiment in Ref.~\cite{parres}  was performed with the following values of dimensionless parameters:
\begin{equation}
p=15,\quad  q=10,\quad \lambda=0.021.
\label{eq:exppar}
\end{equation}
 In Fig. \ref{fig:gp}, we plot the resulting time-dependent axial and radial condensate widths $\rho_{\mathrm{rms}}(t)$ and $z_{\mathrm{rms}}(t)$ calculated as root mean square values
\begin{eqnarray}
\rho_{\mathrm{rms}}(t) =\sqrt{2\pi \int_{-\infty}^{\infty} \! dz\,\int_0^{\infty} \! \rho \, d\rho\, |\psi(\rho,z,t)|^2\,\rho^2}\,,\\
z_{\mathrm{rms}}(t) =\sqrt{2\pi \int_{-\infty}^{\infty} \! dz\, \int_0^{\infty} \! \rho\, d\rho\,  |\psi(\rho,z,t)|^2\, z^2 }\,,
\end{eqnarray}
and compare them with numerical solutions of Eqs.~(\ref{eq:var1}) and (\ref{eq:var2}).
We assume that initially the condensate is in the ground state. In the variational description, this translates into initial conditions $u_{\rho}(0)=u_{\rho0}$, $\dot{u}_{\rho}(0)=0$, $u_{z}(0)=u_{z0}$, $\dot{u}_{z}(0)=0$, where $u_{\rho0}$ and  $u_{z0}$ are time-independent solutions of Eqs.~(\ref{eq:var1}) and (\ref{eq:var2}), while in GP simulations we reach the ground state by performing an imaginary-time propagation \cite{numericsAdhikari}. For solving the GP equation (\ref{eq:gp}), we use the split-step Crank-Nicolson method \cite{numericsAdhikari}. It is evident that we have a good qualitative agreement between the two approaches. 

The main result obtained previously by using the Gaussian approximation is an analytical estimate for the frequencies of the low-lying collective modes \cite{variational, variational2}. In this paper we consider excitations induced by a modulation of the interaction strength, and focus on the properties of the quadrupole and breathing mode. We assume that the external trap is stationary, thus preventing excitations of the dipole (Kohn) mode, corresponding to the center of mass motion. By linearizing the Eqs.~(\ref{eq:var1}) and (\ref{eq:var2}) around the equilibrium widths $u_{\rho 0}$ and $u_{z0}$, frequencies of both the quadrupole $\omega_{Q0}$ and the breathing mode $\omega_{B0}$ were obtained:
\begin{eqnarray}
& &\hspace{-6mm}\omega_{B0,Q0}=\sqrt{2}\left[\left(1+\lambda^2-\frac{p}{4u_{\rho 0}^2 u_{z0}^3}\right)\right.
\nonumber \\ 
 &&\hspace*{-2mm}\pm \left.\sqrt{\left(1-\lambda^2+\frac{p}{4u_{\rho 0}^2 u_{z0}^3}\right)^2+8\left(\frac{p}{4u_{\rho 0}^3 u_{z0}^2}\right)^2} \right]^{\frac{1}{2}}\!.
\label{eq:linres}
\end{eqnarray}
For the repulsive interaction, the quadrupole mode has a lower frequency and is characterized by out-of-phase radial and axial oscillations, while in-phase oscillations correspond to the breathing mode. In the case of the experiment \cite{parres}, Eq.~(\ref{eq:linres}) yields: 
\begin{equation}
\omega_{Q0}=0.035375,\quad \omega_{B0}=2.00002.
\label{eq:expfrequencies}
\end{equation}
We emphasize that, although based on the Gaussian ansatz, the variational approximation reproduces exactly the frequencies of collective modes not only for the weakly interacting BEC but also for the strongly interacting BEC in the Thomas-Fermi regime \cite{stringari, variational}. Therefore, it represents a solid analytical description of BEC dynamics.

\begin{figure*}[!t]
\begin{center}
\includegraphics[width=5.5cm]{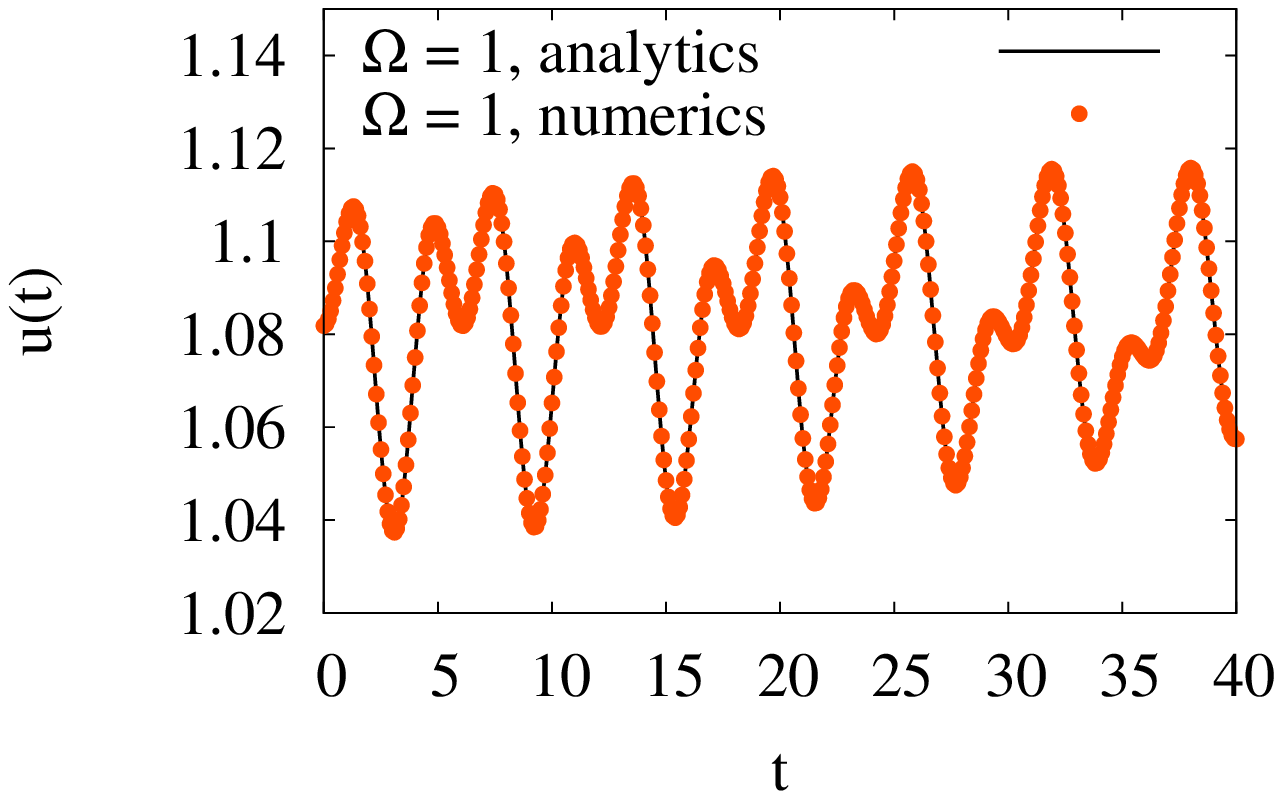}
\includegraphics[width=5.5cm]{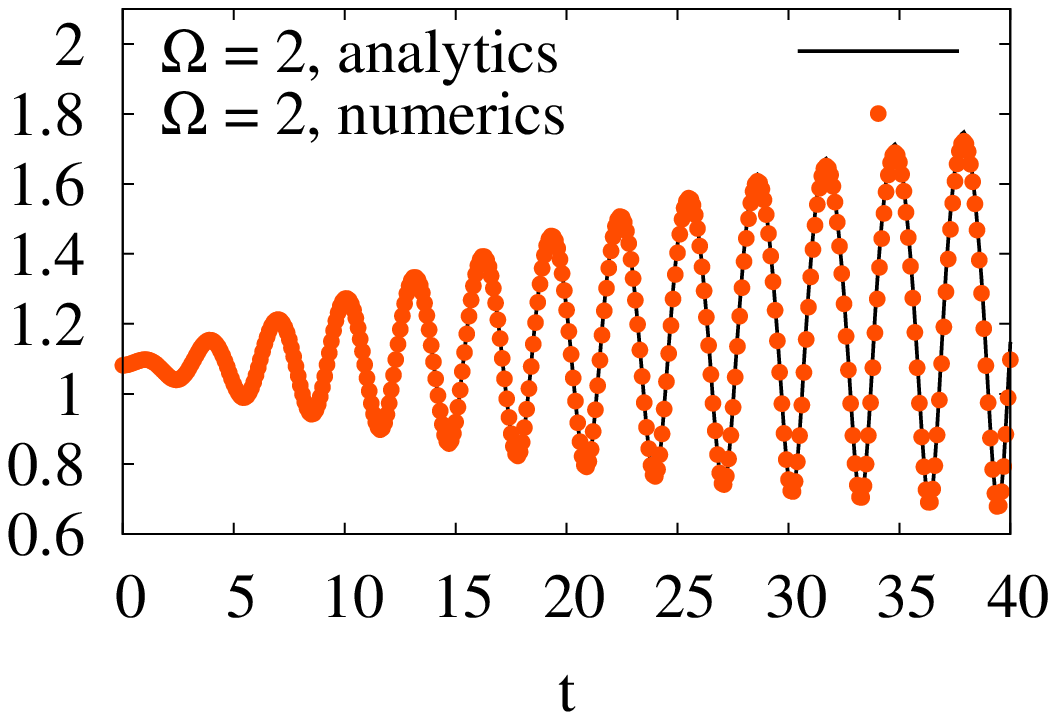}
\includegraphics[width=5.5cm]{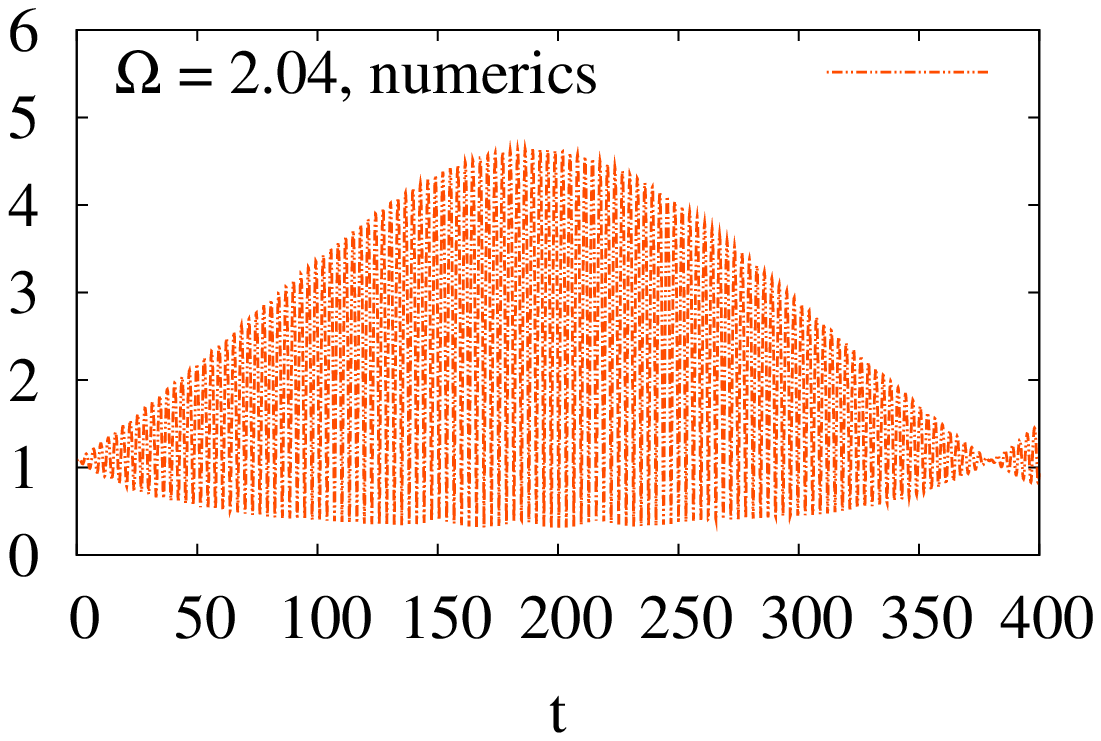}
\includegraphics[width=5.5cm]{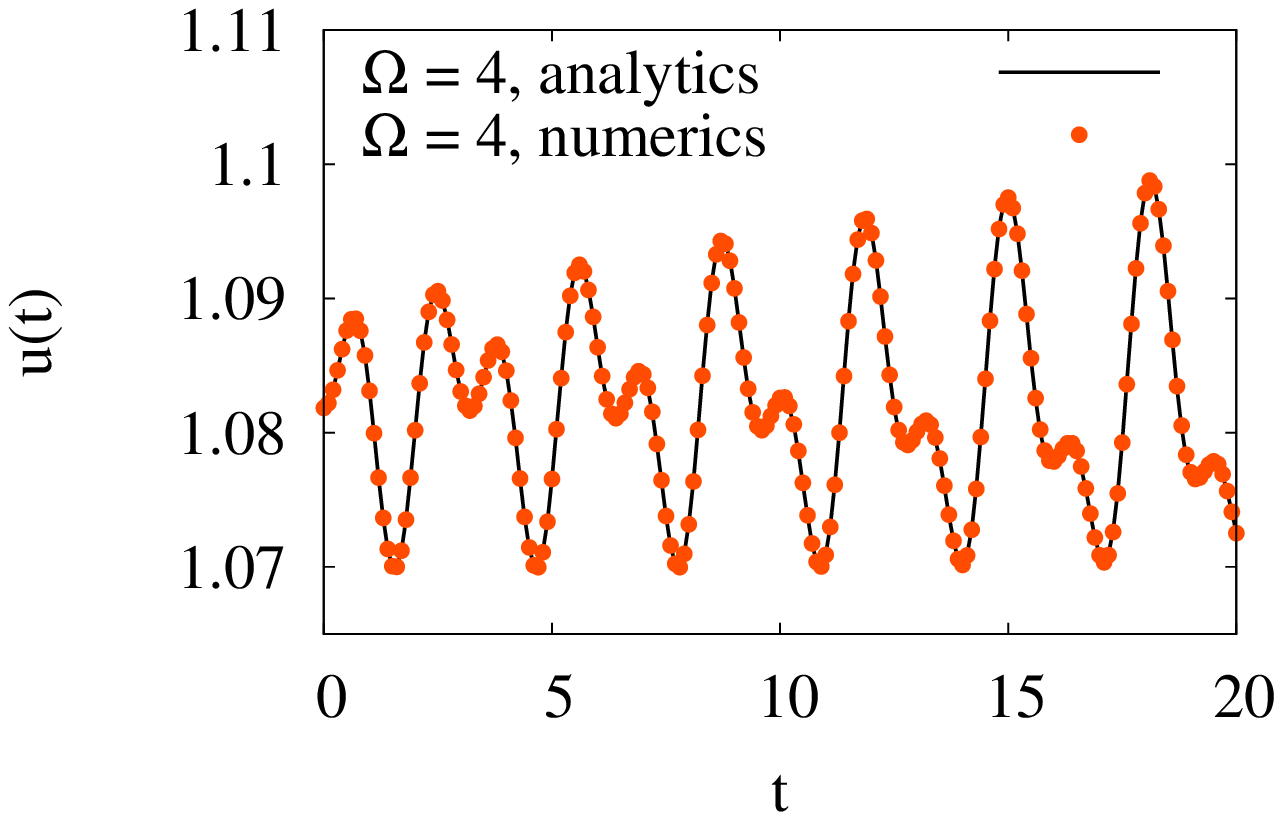}
\includegraphics[width=5.5cm]{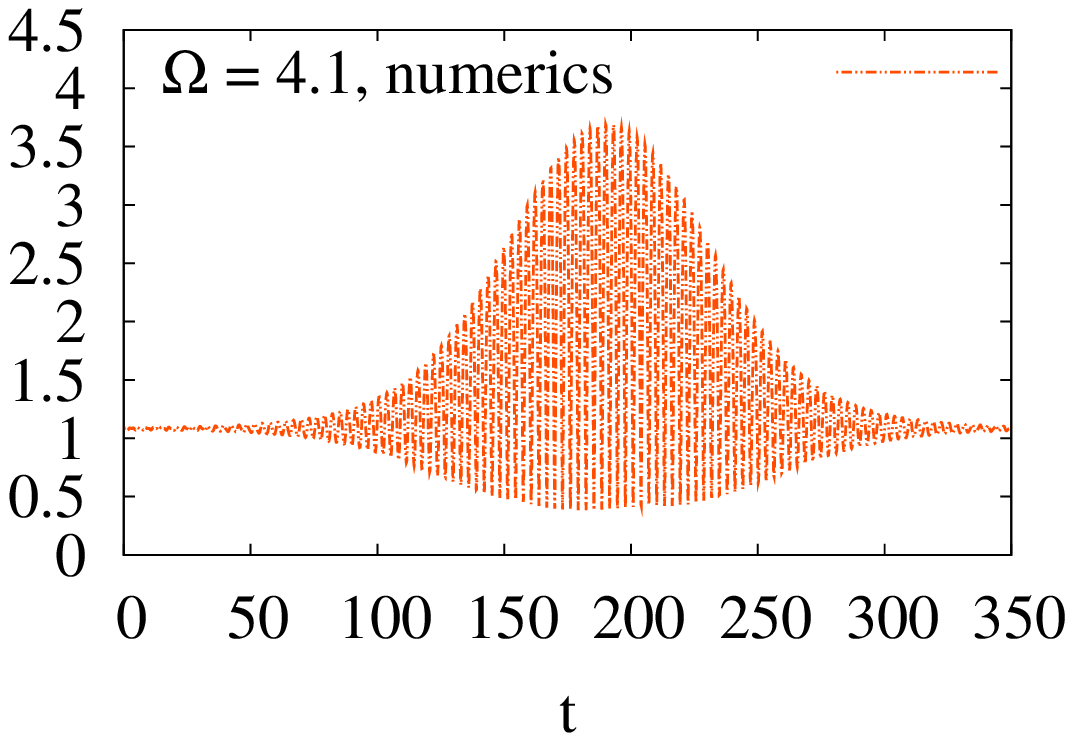}
\includegraphics[width=5.5cm]{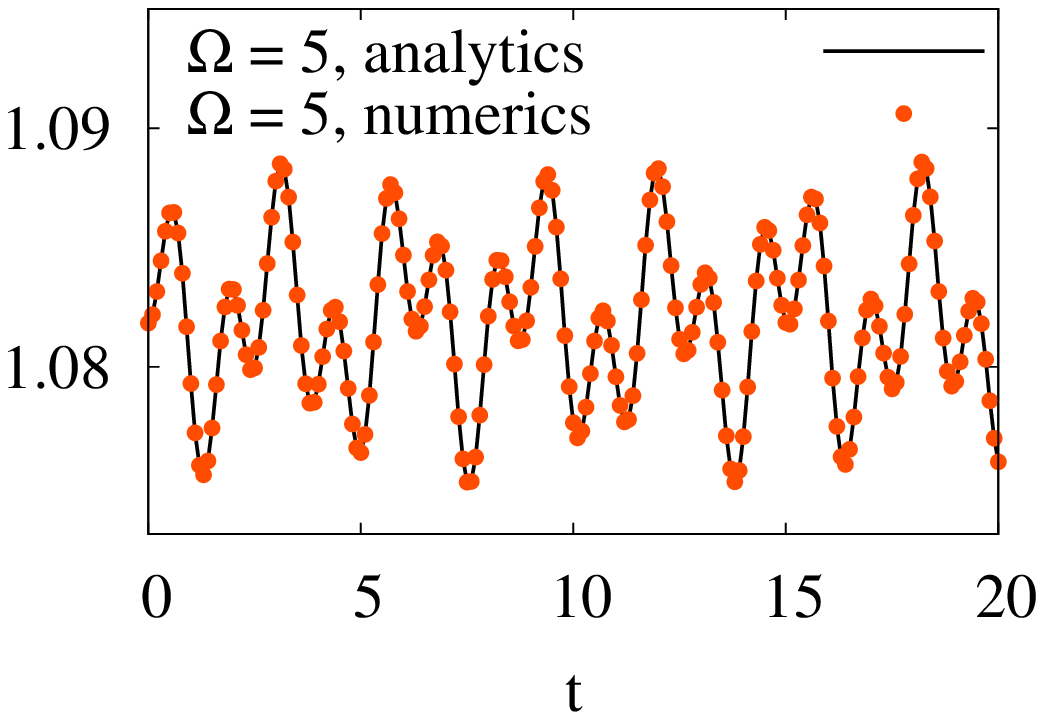}
\caption{(Color online) Condensate dynamics $u(t)$ versus $t$ within the Gaussian approximation for $p=0.4$, $q=0.1$ and several different driving frequencies $\Omega$. We plot the exact numerical solution of Eq.~(\ref{eq:sphvar}). For off-resonant driving frequencies $\Omega$, we also show our analytical third-order perturbative result, as explained in Section III.B.}
\label{fig:sphrt}
\end{center}
\end{figure*}

However, due to the nonlinear form of the underlying GP equation, we expect nonlinearity-induced shifts in the frequencies of low-lying modes with respect to the values in Eq.~(\ref{eq:linres}) calculated using the linear stability analysis. In particular, our goal is to describe collective modes induced by the harmonic modulation of the interaction strength in Eq.~(\ref{eq:modint}). In the case of a close matching of the driving frequency $\Omega$ and one of the BEC eigenmodes we expect resonances, i.e. large amplitude oscillations. Here, the role of the nonlinear terms becomes crucial and nonlinear phenomena become visible, as we discuss in the next section.

\section{Spherically symmetric BEC}

Using a simple symmetry-based reasoning, we conclude that a harmonic modulation of interaction strength in the case of a spherically symmetric BEC, i.e. $\lambda=1$, leads to the excitation of the breathing mode only, so that $u_{\rho}(t)=u_z(t)\equiv u(t)$. This fact simplifies numerical and analytical calculations
and this is why we first consider this case before we embark to the study of a more complex axially symmetric BEC. 

Thus, the system of ordinary differential Eqs.~(\ref{eq:var1}) and (\ref{eq:var2}) reduces to a single equation:
\begin{equation}
 \ddot{u}(t)+u(t)-\frac{1}{u(t)^3}-\frac{p(t)}{u(t)^4}=0 \, .
\label{eq:sphvar}
\end{equation}
The equilibrium condensate width $u_0$ satisfies
\begin{equation}
u_0-\frac{1}{u_0^3}-\frac{p}{u_0^4}=0\, ,
\label{eq:u0}
\end{equation}
and a linear stability analysis yields the breathing mode frequency
\begin{equation}
\omega_0=\sqrt{1 + \frac{3}{u_0^4} + \frac{4 p}{u_0^5}}\, .
\label{eq:omega0}
\end{equation}
Note that the above result for the breathing mode can be also obtained from Eq.~(\ref{eq:linres}) if we set $\lambda=1$,  $u_{\rho 0}=u_{z 0}\equiv u_0$, and take into account Eq.~(\ref{eq:u0}).

\subsection{Numerical simulations}

The main feature of the modulation induced dynamics is that it strongly depends on the value of the driving frequency $\Omega$. To illustrate this, we set $p=0.4$, $q=0.1$ and solve Eq.~(\ref{eq:sphvar}) for different values of $\Omega$. From the linear response theory, we have $u_0=1.08183$, $\omega_0=2.06638$ and  we assume that the condensate is initially in equilibrium, i.e. $u(0)=u_0, \dot u(0)=0$. Numerical results are plotted in Fig.~\ref{fig:sphrt}. Large amplitude oscillations and beating phenomena are present for both $\Omega\approx\omega_0$ and for $\Omega\approx 2 \omega_0$. 

\begin{figure}[!t]
\begin{center}
\includegraphics[width=8cm]{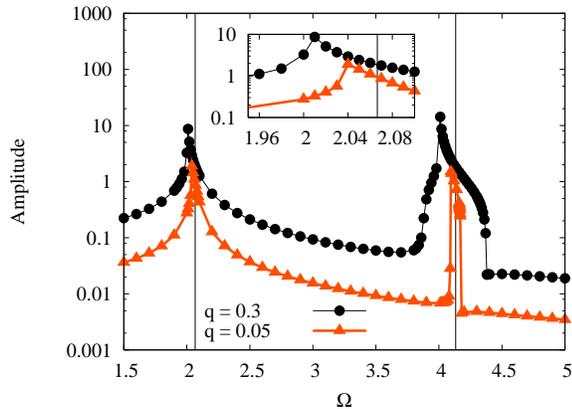}
\caption{(Color online) Oscillation amplitude $(u_{\mathrm{max}}-u_{\mathrm{min}})/2$ versus driving frequency $\Omega$ for $p=0.4$. In the inset, we zoom to the first peak to emphasize that the shape and value of a resonance occur at a driving frequency $\Omega$ which differs from $\omega_0 $ depends on the modulation amplitude $q$. The solid vertical lines correspond to $\omega_0$ and $2\omega_0$.}
\label{fig:response}
\end{center}
\end{figure}

The phenomenology based on Eq.~(\ref{eq:sphvar}) is more systematically shown in Fig.~\ref{fig:response} where we plot the oscillation amplitude defined as $(u_{\mathrm{max}}-u_{\mathrm{min}})/2$ versus the driving frequency $\Omega$. A resonant behavior becomes apparent for both $\Omega\approx\omega_0$ and $\Omega\approx2\omega_0$. In the same figure we also show the expected positions of resonances calculated using the linear stability analysis. Clearly, the prominent peaks exhibit shifts with respect to the solid vertical lines representing $\omega_0$ and $2\omega_0$. As expected, a stronger modulation amplitude leads to a larger frequency shift, as can be seen from the inset.

\begin{figure}[!t]
\begin{center}
\includegraphics[width=6cm]{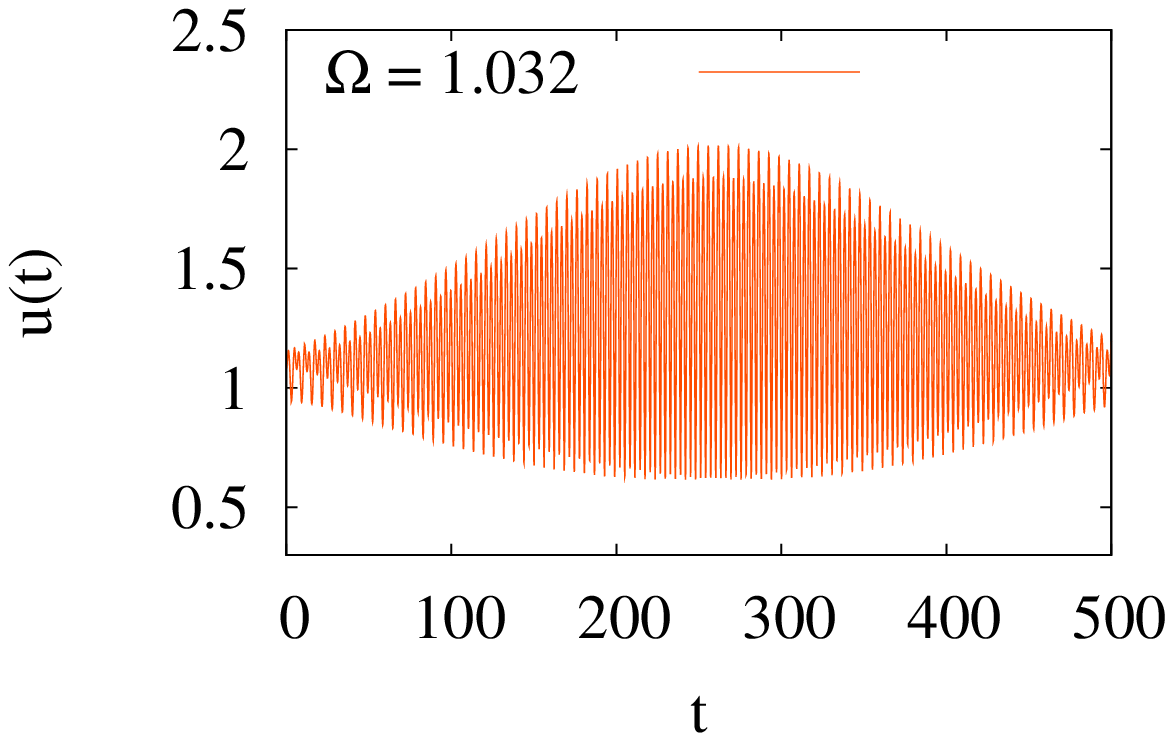}
\includegraphics[width=6cm]{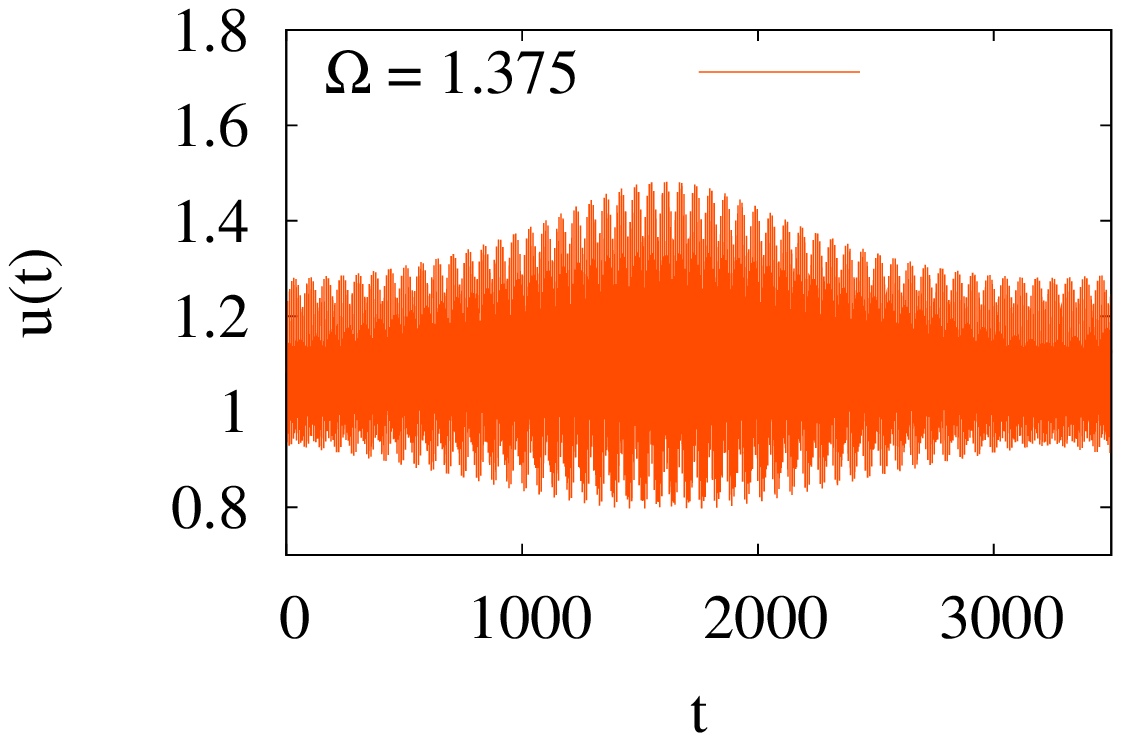}
\caption{(Color online) Exact numerical solution of Eq.~(\ref{eq:sphvar}) for the condensate width $u(t)$ versus $t$ for $p=0.4$, $q=0.3$, corresponding to $\omega_0=2.06638$. We observe large amplitude oscillations for $\Omega \approx \omega_0/2$ in the top panel, while in the bottom panel the ``resonant'' behavior is present for $\Omega\approx2\omega_0/3$.}
\label{fig:sphrthigherharmonics}
\end{center}
\end{figure}

The curves presented in Fig.~\ref{fig:response} are obtained by an equidistant sampling of the external driving frequency $\Omega$. 
In addition to the expected resonances close to $\omega_0$ and $2\omega_0$, a more thorough exploration of solutions of the variational equation (\ref{eq:sphvar})
shows that other ``resonances'' are present such as, e.~g.~ at $\Omega \approx \omega_0/2 $ and $\Omega \approx 2\omega_0/3 $. This is further demonstrated in Fig.~\ref{fig:sphrthigherharmonics}. These ``resonances'' are harder to observe numerically, since it is necessary to perform a fine tuning of the external frequency. 
However, they clearly demonstrate nonlinear BEC properties and an experimental observation of these phenomena is certainly of high interest. We note that the observed resonance pattern of the form $\Omega\approx2\omega_0/n$ (where $n$ is an integer) arises also in the case of a parametrically driven system described by the Mathieu equation, for instance, in the context of the Paul trap \cite{rmpPaultrap}.

\begin{figure}[!ht]
 \begin{center}
\includegraphics[width=8cm]{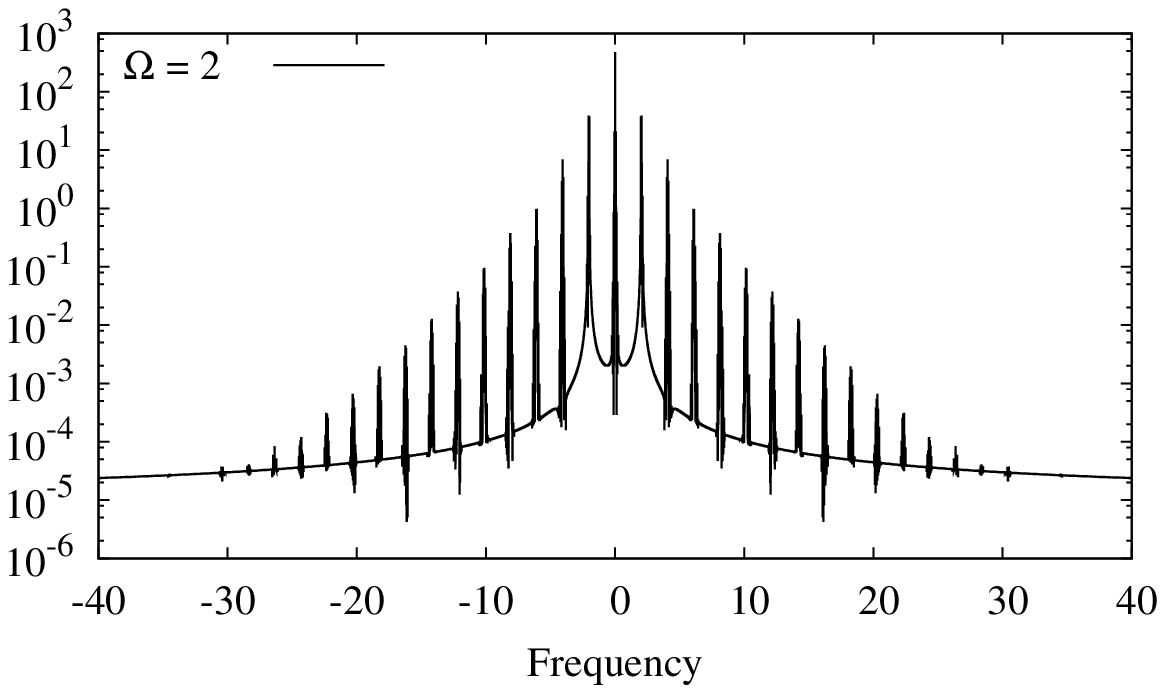}
\includegraphics[width=4cm]{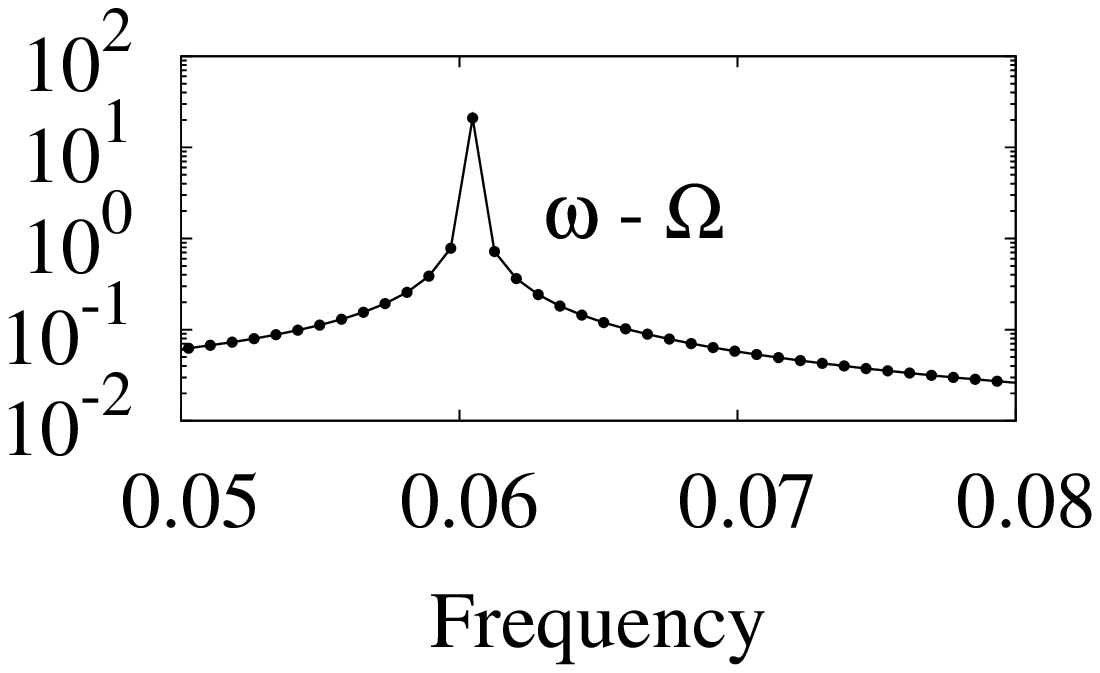}
\includegraphics[width=4cm]{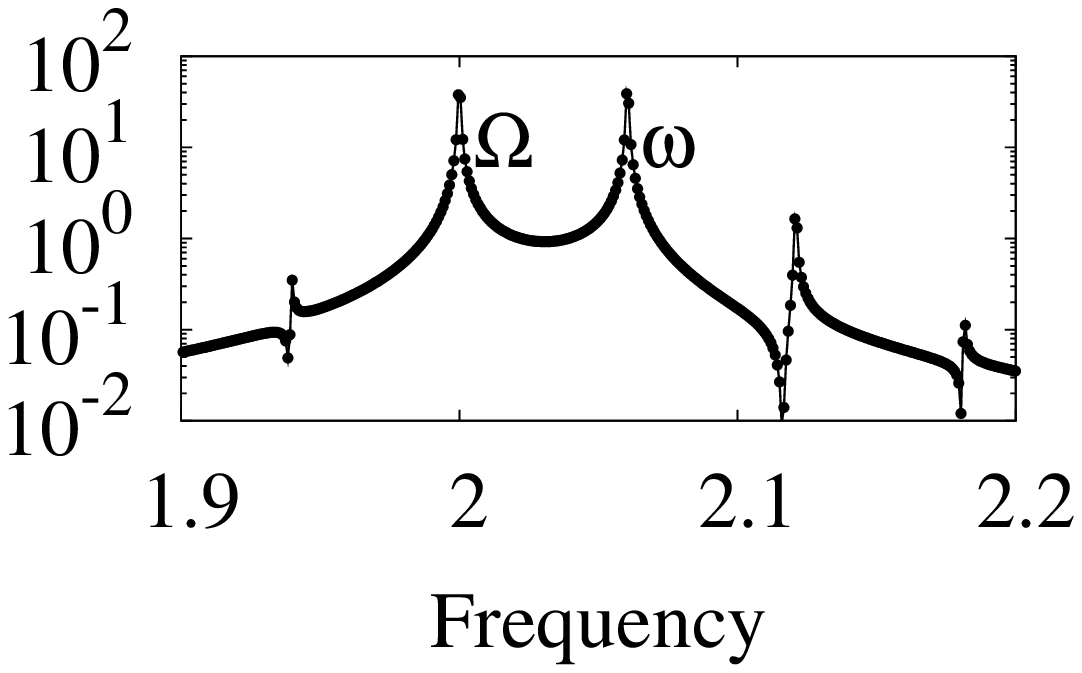}
\includegraphics[width=4cm]{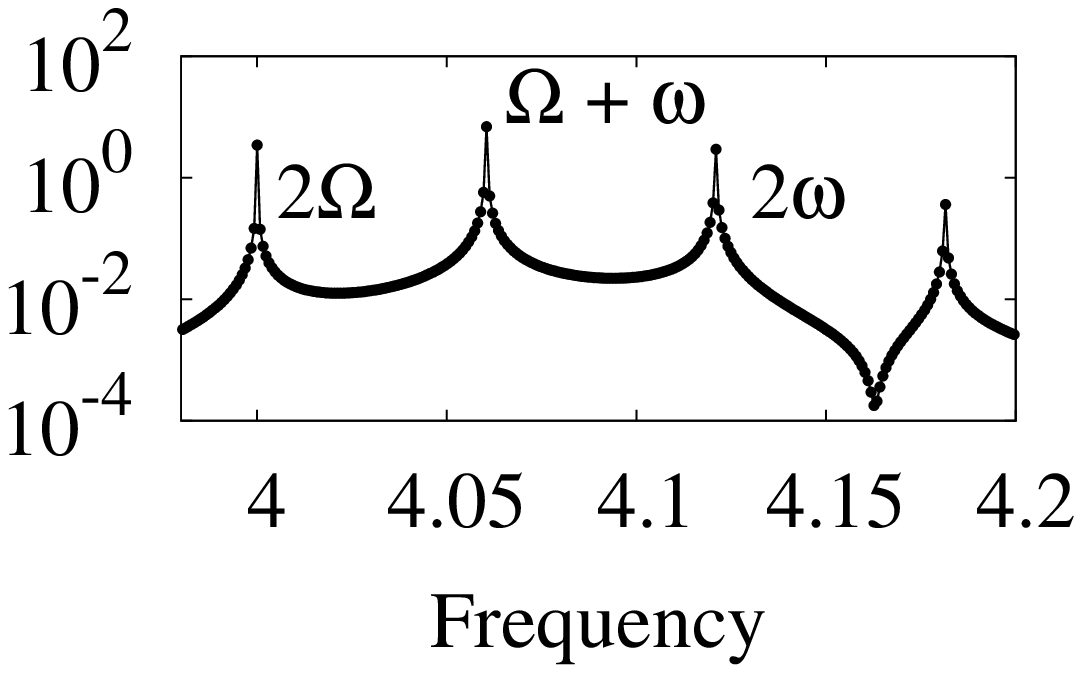}
\includegraphics[width=4cm]{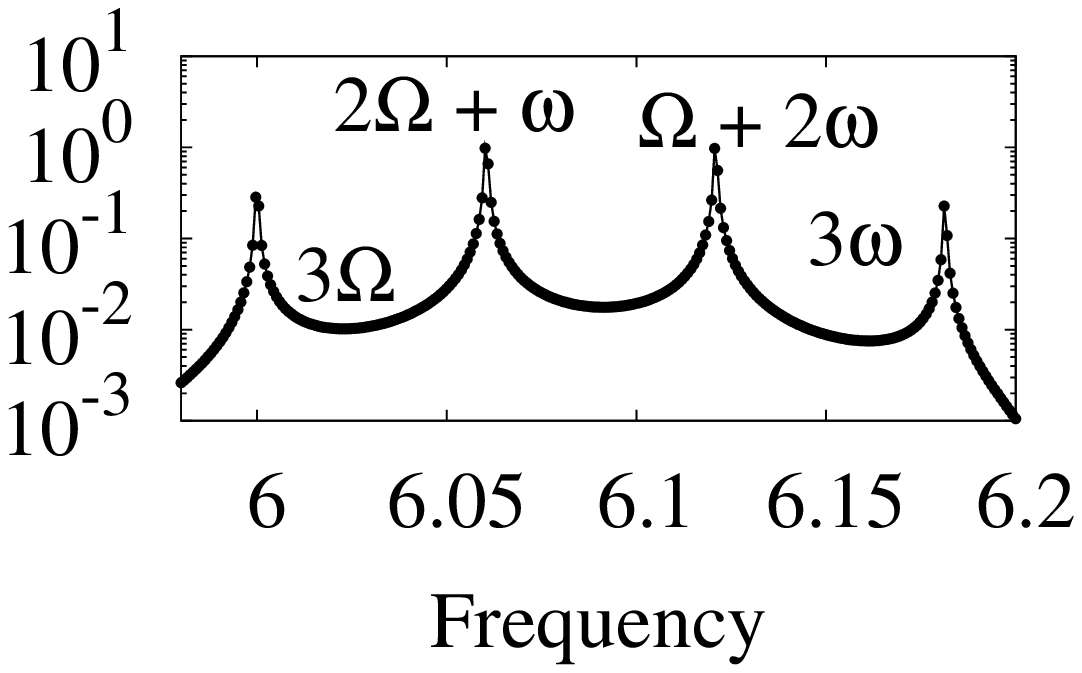}
\caption{ Fourier transform of $u(t)$ for $p=0.4$, $q=0.1$, and
$\Omega=2$. First plot presents the complete spectrum on a semi-log scale, while the subsequent plots focus on regions of interest in the spectrum.}
\label{fig:sphft}
\end{center}
\end{figure}

To examine such excited modes directly, we look at the Fourier transform of the condensate width $u(t)$.
To this end, we numerically solve Eq.~(\ref{eq:sphvar}) and find the Fourier transform of its solution using the Mathematica software package \cite{Mathematica}. An example of such an  excitation spectrum for $p=0.4$, $q=0.1$, and $\Omega=2$ is given in Fig.~\ref{fig:sphft}. The spectrum contains two prominent modes  - a  breathing mode of frequency $\omega$ (close, but not equal to $\omega_0$) and a mode that corresponds to the driving frequency $\Omega$, along with many higher-order harmonics which are of the general form $m \Omega+n \omega$, where $m$ and $n$ are integers.

\begin{figure}[!t]
\begin{center}
\includegraphics[width=4cm]{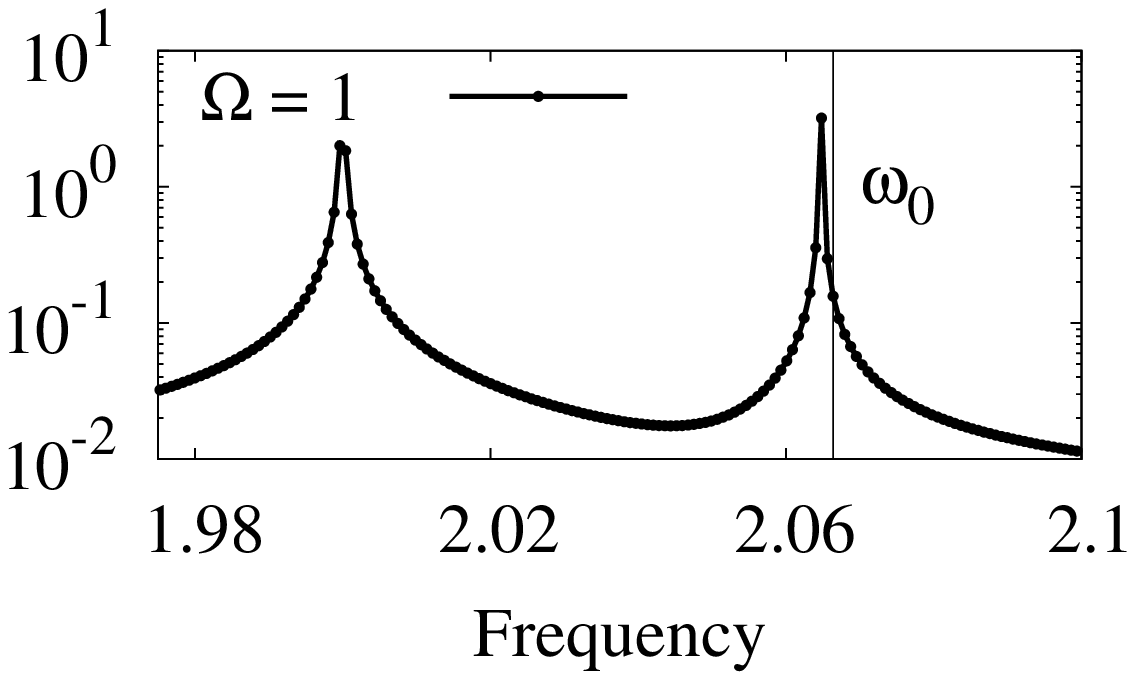}
\includegraphics[width=4cm]{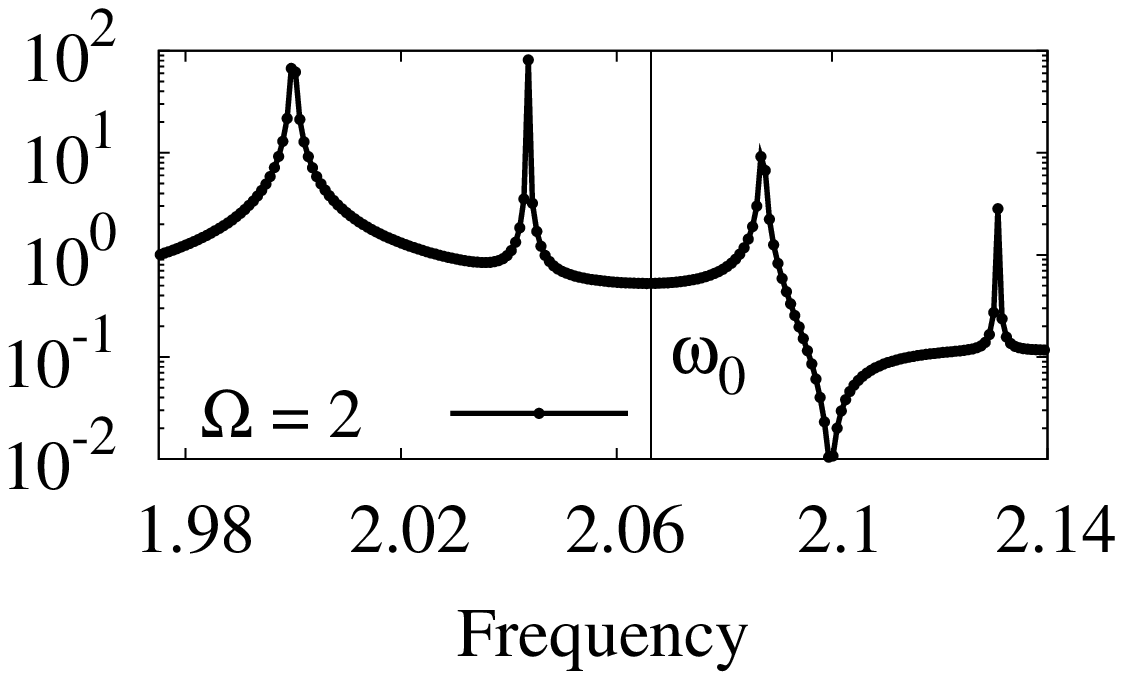}
\caption{ Parts of the Fourier spectra for $p=0.4$, $q=0.2$, and two different driving frequencies:
$\Omega=1$ (left) and $\Omega=2$ (right).
Position of a linear response result $\omega_0$ is given by a vertical solid line.
}
\label{fig:juxtaposition}
\end{center}
\end{figure}

In Fig.~\ref{fig:juxtaposition} we juxtapose two zoomed Fourier spectra for two different driving frequencies for $p=0.4$ and $q=0.2$. On the left plot, we show zoomed spectrum for $\Omega=1$.
The vertical solid line corresponds to $\omega_0$ and we find the peak in the spectrum that lies almost precisely at this position. On the contrary, from the right plot of Fig.~\ref{fig:juxtaposition}, that corresponds almost to the resonant excitation $\Omega=2$, we see that the prominent peak is displaced from the vertical line.
This is the most clear-cut illustration of the shifted eigenfrequency arising due to the nonlinearity of the underlying dynamical equations. Our objective is to develop an analytical approach capable of taking into account these nonlinear effects. 
 
\subsection{Poincar\'e-Lindstedt method}
In its essence, our analytical approach represents the standard Poincar\'e-Lindstedt method \cite{book1, book2, book3, poinlind}. 
Linearizing the variational equation (\ref{eq:sphvar}) around the
time-independent solution $u_0$ for vanishing driving $q=0$, we obtain the
zeroth-order approximation for the collective mode $\omega=\omega_0$ expressed by Eq.~(\ref{eq:omega0}).
To calculate the collective mode to higher orders, we explicitely introduce the sought-after eigenfrequency $\omega$ into the calculation by rescaling the time from $t$ to $s = \omega t$, yielding the equation:
\begin{equation}
 \omega^2\, \ddot{ u}(s) + u(s) - \frac{1}{u(s)^3} - \frac{p}{u(s)^4} - 
 \frac{q}{u(s)^4}\, \cos\frac{\Omega s}{\omega}=0\, .
\label{eq:rescaled}
\end{equation}
In the next step, we assume the following perturbative expansions in the modulation amplitude $q$:
\begin{eqnarray}
u(s) &=& u_0 + q\, u_1(s) + q^2\, u_2(s) + q^3\, u_3(s)+\ldots\, ,\label{eq:expansions1}\\
\omega &=& \omega_0 + q\, \omega_1 + q^2\, \omega_2 + q^3\, \omega_3+\ldots\, ,
\label{eq:expansions2} 
\end{eqnarray}
where we expand $\omega$ around $\omega_0$ and introduce frequency shifts $\omega_1 $, $\omega_2$, \ldots for each order in the expansion in $q$.
By inserting the above expansions into the Eq.~(\ref{eq:rescaled}) and collecting terms of the same order in $q$, we obtain a hierarchical system of linear differential equations. Up to the third order,
we find:
\begin{eqnarray*}
&&\hspace{-5mm}\omega_0^2 \ddot{u}_1(s)+\omega_0^2 u_1(s)=\frac{1}{u_0^4}\cos\frac{\Omega
s}{\omega}\, ,\\
&&\hspace{-5mm}\omega_0^2 \ddot{u}_2(s)+\omega_0^2 u_2(s)=\\
&-&2 \omega_0 \omega_1 \ddot{ u}_1(s)-\frac{4}{u_0^5} u_1(s) \cos\frac{\Omega
s}{\omega} +\alpha u_1(s)^2,\\
&&\hspace{-5mm}\omega_0^2 \ddot{u}_3(s)+\omega_0^2 u_3(s)=\\
&-&2 \omega_0 \omega_2 \ddot{u}_1(s)-2\beta u_1(s)^3+2\alpha u_1(s) u_2(s)
-\omega_1^2 \ddot{u}_1(s)\\
&+& \frac{10}{ u_0^6}\, u_1(s)^2\, \cos\frac{\Omega s}{\omega} -\frac{4}{
u_0^5}\, u_2(s)\, \cos\frac{\Omega s}{\omega}-2 \omega_0\, \omega_1\,
\ddot{u}_2(s),\\
\end{eqnarray*}
where $\alpha=10\,p/u_0^6+6/u_0^5$ and $\beta=10\,p/u_0^7+5/u_0^6$.

These equations disentangle in a natural way - we solve the first one for $u_1(s)$ and use that solution to solve the second one for $u_2(s)$ and so on. At the $n$-th level of the perturbative expansion ($n \geq 1$) we use the  initial conditions $u_n(0)=0, \dot u_n(0)=0$. As is well known, the presence of the term $\cos s$ on the right-hand side of some of the previous equations would yield a solution that contains the secular term $s \sin s$. Such a secular term grows linearly in time, which makes it the dominant term in the expansion (\ref{eq:expansions1}) that otherwise contains only periodic functions in $s$. In order to ensure a regular behavior of the perturbative expansion, the respective frequency shifts $\omega_1$, $\omega_2, \ldots$  are determined by imposing the cancellation of secular terms.

This analytical procedure is implemented up to the third order in the modulation amplitude $q$ by using the software package Mathematica \cite{Mathematica}. Although the calculation is straightforward, it easily becomes tedious for higher orders of perturbation theory. Note that it is necessary to perform calculation to at least the third order since it turns out to be  the lowest-order solution where secular terms appear and where nontrivial frequency shift is obtained. We solve explicitely for $u_1(s)$, $u_2(s)$, and $u_3(s)$ and show an excellent agreement of our analytical solutions with a respective numerical solution of Eq.~(\ref{eq:sphvar}) in Fig.~\ref{fig:sphrt}. From the first-order solution $u_1(t)$ we read off  only the two basic modes $\omega_0$ and $\Omega$, while the second-order harmonics $2 \omega_0$, $\omega_0-\Omega$, $\omega_0+\Omega$ and $2\Omega$ appear in $u_2(t)$. In the third order of perturbation theory, higher-order harmonics $\omega-2\Omega$, $2\omega-\Omega$, $2\omega+\Omega$, $\omega+2\Omega$, $3\omega$, and $3\Omega$ are also present. Concerning the cancellation of secular terms, the first-order correction $\omega_1$ vanishes, leading to a frequency shift which is  quadratic in $q$:
\begin{equation}
\omega=\omega_0+\frac{q^2}{12 u_0^{20} \omega_0^3}\,\frac{\mathrm{P}(\Omega)}{(\Omega^2-\omega_0^2)^2\, (\Omega^2-4\omega_0^2)} +\ldots\, , 
\label{eq:sphresult}
\end{equation}
where the polynomial $\mathrm{P}(\Omega)$ is given by
\begin{eqnarray}
\mathrm{P}(\Omega)&=&\Omega^4 \left[-240 p u_0^5 + 
    36 u_0^6 (-4 + 3 u_0^4 \omega_0^2)\right]\nonumber\\
      &+& \Omega^2 \left[-1100 p^2 - 
    30 p u_0 (44 - 65 u_0^4 \omega_0^2)\right.\nonumber\\ 
      &+& \left.
    9 u_0^2 (-44 + 127 u_0^4 \omega_0^2 - 44 u_0^8 \omega_0^4)\right]\nonumber\\
     &+& 5600 p^2 \omega_0^2 - 
 840 p u_0 \omega_0^2 (-8 + 3 u_0^4 \omega_0^2) \nonumber\\
&+& 
 36 u_0^2 \omega_0^2 (56 - 39 u_0^4 \omega_0^2 + 8 u_0^8 \omega_0^4).
\end{eqnarray}
Mathematica notebook which implements this analytical calculation is avaliable at our web site \cite{scl}.

\begin{figure}[!t]
\begin{center}
\includegraphics[width=8cm]{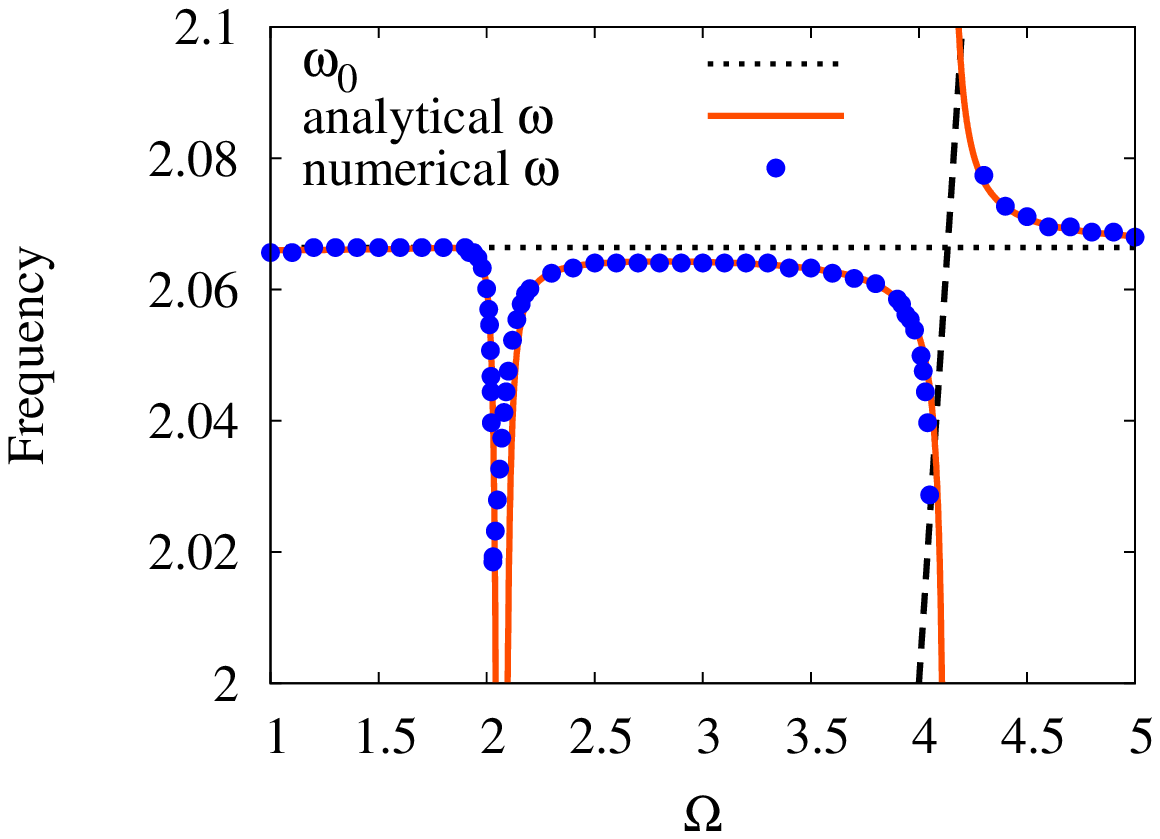}
\includegraphics[width=8cm]{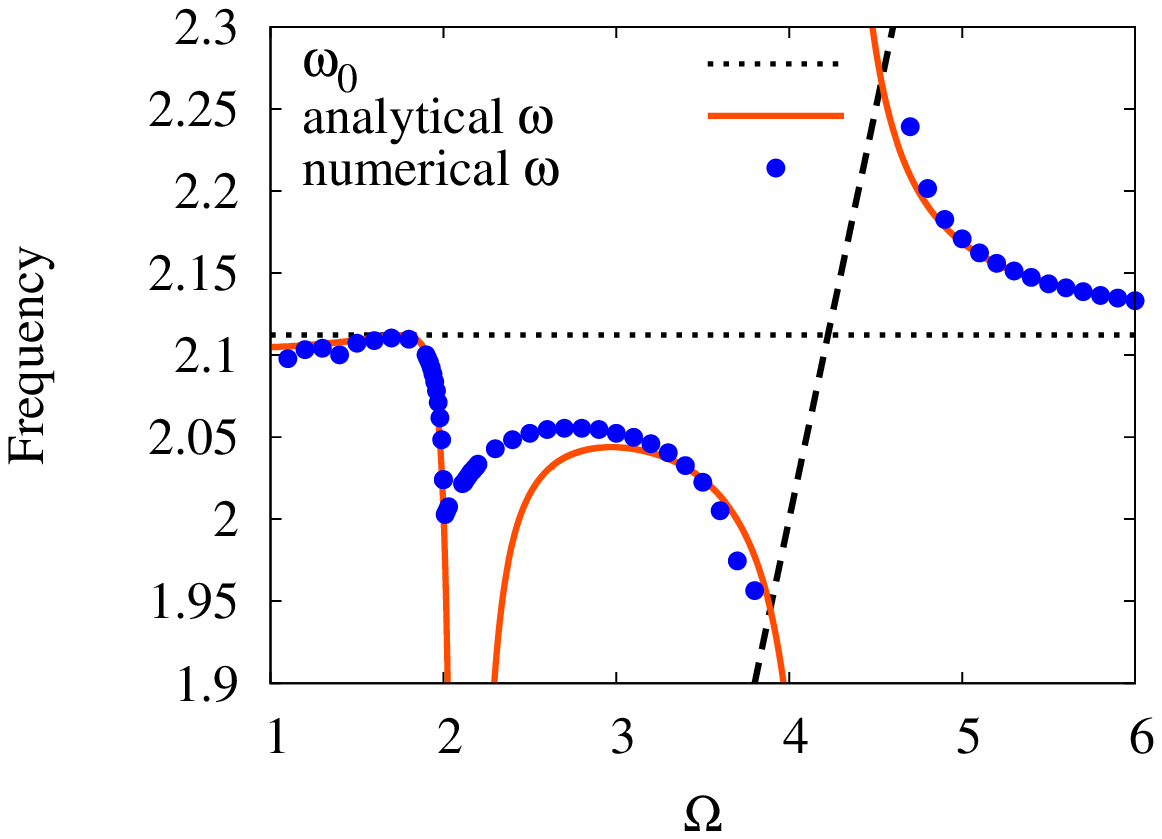}
\caption{(Color online) Frequency of the breathing mode versus the driving frequency $\Omega$ for $ p=0.4$ and $q=0.1$ (upper plot), and $p=1$ and $q=0.8$ (lower plot). The dashed line represents $\Omega/2$ and is given to guide the eye.}
\label{fig:sphresult}
\end{center}
\end{figure}

\subsection{Discussion}
The result given by Eq.~(\ref{eq:sphresult}) is the main achievement of our analytical analysis. It is obtained within a perturbative approach up to the second order in $q$ and it describes the breathing mode frequency dependence on $\Omega$ and $q$ as a result of nonlinear effects. Due to the underlying perturbative expansion, we do not expect Eq.~(\ref{eq:sphresult}) to be meaningful at the precise position of the resonances. However, by comparison with numerical results based on the variational equation, we find that Eq.~(\ref{eq:sphresult}) represents a reasonable approximation even close to the resonant region. 

To illustrate this, in Fig.~\ref{fig:sphresult} we show two such comparisons. In the upper panel we consider the parameter set $p=0.4$ and $q=0.1$ and observe significant frequency shifts only in the narrow resonant regions. We notice an excellent agreement of numerical values with the analytical result given by Eq.~(\ref{eq:sphresult}). In the lower panel we consider the parameter set $p=1$ and $q=0.8$ with much stronger modulation amplitude. In this case we observe significant frequency shifts for the broader range of modulation frequencies $\Omega$. In spite of a strong modulation, we still see a qualitatively good agreement of numerical results with the analytical prediction given by Eq.~(\ref{eq:sphresult}). In principle, better agreement can be achieved using higher-order perturbative approximation. The dashed line on both figures represents $\Omega/2$, given as a  guide to the eye. It also serves as a crude description of what we observe numerically in the range $\Omega\approx2\omega_0$.

The presence of two poles at $\Omega=\omega_0$ and $\Omega=2\omega_0$ in Eq.~(\ref{eq:sphresult}) implies the possible existence of real resonances in the BEC with a  harmonically modulated interaction.
A perturbative expansion to higher orders would probably introduce some additional poles,
responsible for higher-order ``resonant'' behavior observed at $\Omega\approx2\omega_0/n$, ($n\geq3$).
Still, the poles seem to be only an artefact of our approximative perturbative scheme, not present in the exact description. For example, a simple resummation performed by using the second-order perturbative result removes these effects, although this is only an ad-hoc approximation. We stress that this issue concerning the true resonant behavior can not be settled either by relying on a numerical calculation due to inherent numerical artefacts related to finite numerical precision and finite computational time. To resolve it, one should rely on an analytical consideration along the lines of Ref.~\cite{timedeptrap2} or use some analytical tool applicable at resonances, such as the resonant Bogoliubov-Mitropolsky method \cite{book3}. However, this is out of scope of the present paper.

\begin{figure}[!b]
\begin{center}
\hspace{-1cm}
\includegraphics[width=8cm]{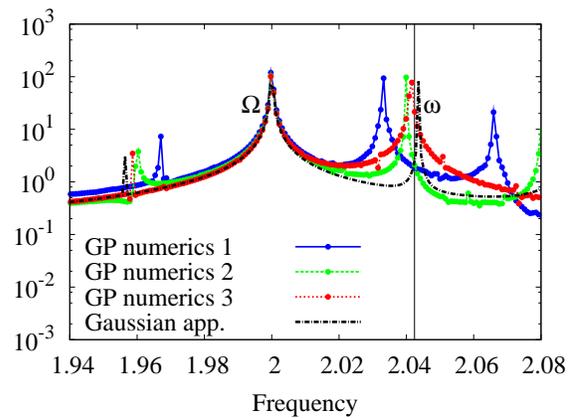}
\end{center}
\caption{(Color online) Part of the Fourier spectrum of the time-dependent condensate width for $p=0.4$, $q=0.2$, $\Omega = 2$. For numerical solution of GP equation we use several discretization schemes: GP numerics 1 (time step $\varepsilon= 10^{-3}$, spacing $h=4 \times 10^{-2}$), GP numerics 2 ($\varepsilon= 5\times 10^{-4} $, $h= 2\times 10^{-2} $), GP numerics 3 ($\varepsilon= 5\times 10^{-5} $, $h= 5\times 10^{-3} $). For comparison we also show the corresponding spectrum obtained from the Gaussian approximation (dotted-dashed line) and analytical result (\ref{eq:sphresult}) for the position of breathing mode (solid vertical line).}
\label{fig:gpcomparison}
\end{figure}

In addition to comparison of our analytical results with numerical solutions based on the Gaussian variational approximation, we present a comparison with the full numerical solution of the GP equation. In order to be able to preform Fourier analysis with good resolution, it is necessary to obtain an accurate solution for long evolution times. We do this by using the split-step method in combination with the semi-implicit Crank-Nicolson method \cite{numericsAdhikari}. As we refine the GP numerics by using finer space and time discretization parameters our numerical results become stable as shown in Fig.~\ref{fig:gpcomparison}. From the same figure, we observe quantitatively good agreement between GP numerics and Gaussian approximation, reflected in close values obtained  for the breathing mode frequency. In addition, numerical values for the breathing mode approach closely the analytical result of Eq.~(\ref{eq:sphresult}), shown by a solid vertical line in Fig.~\ref{fig:gpcomparison}.

It is well known that for a corresponding two-dimen\-sional axially-symmetric system with a constant interaction and trapping frequency, the breathing mode oscillations can be described by an exact linear equation \cite{2dpitaevskii1, 2dpitaevskii2}. However, in the case of a time-dependent trapping frequency, the exact equation of motion is nonlinear \cite{timedeptrap2}. To the best of our knowledge, for a time-dependent interaction strength the corresponding exact equation does not exist in the literature, but one can reasonably expect that nonlinear effects will remain in such systems, due to the inherent time dependence of the interaction.

\section{Axially symmetric BEC}

\begin{figure*}[!hbt]
\begin{center}
\includegraphics[width=8cm]{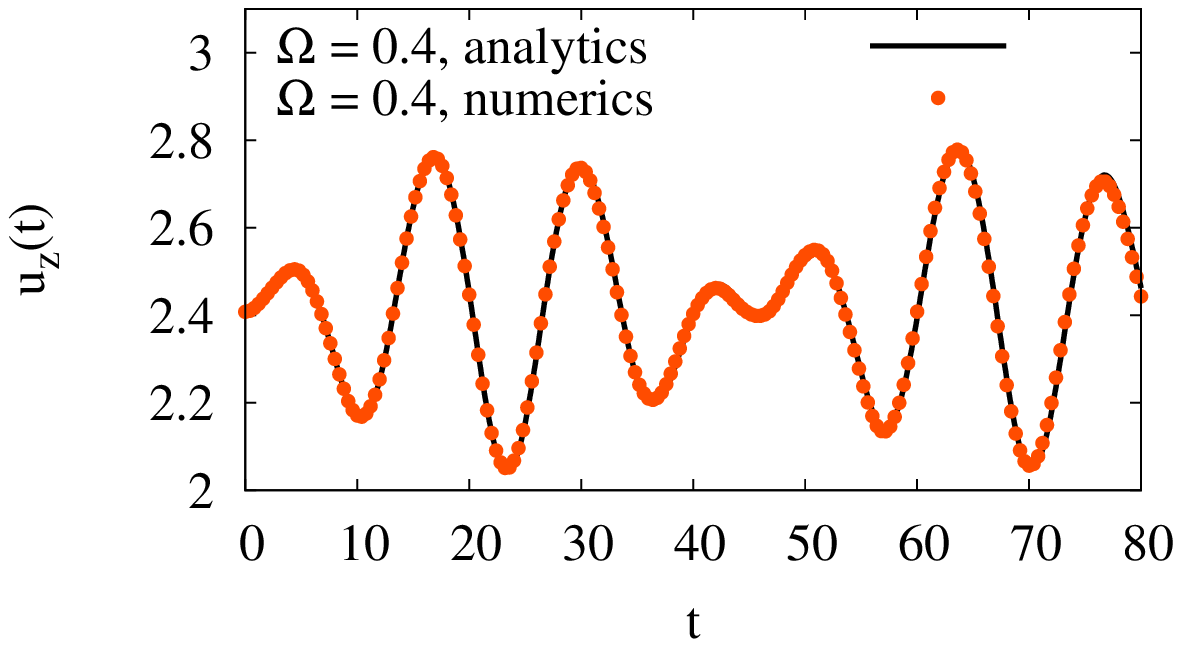}
\includegraphics[width=8cm]{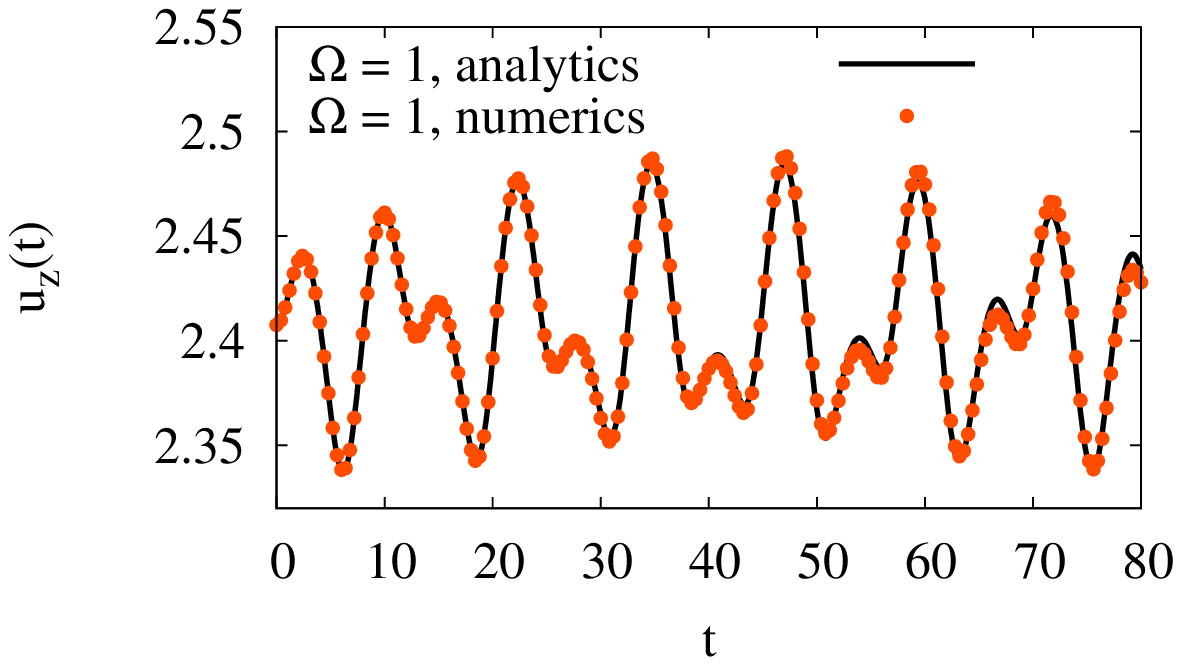}
\includegraphics[width=8cm]{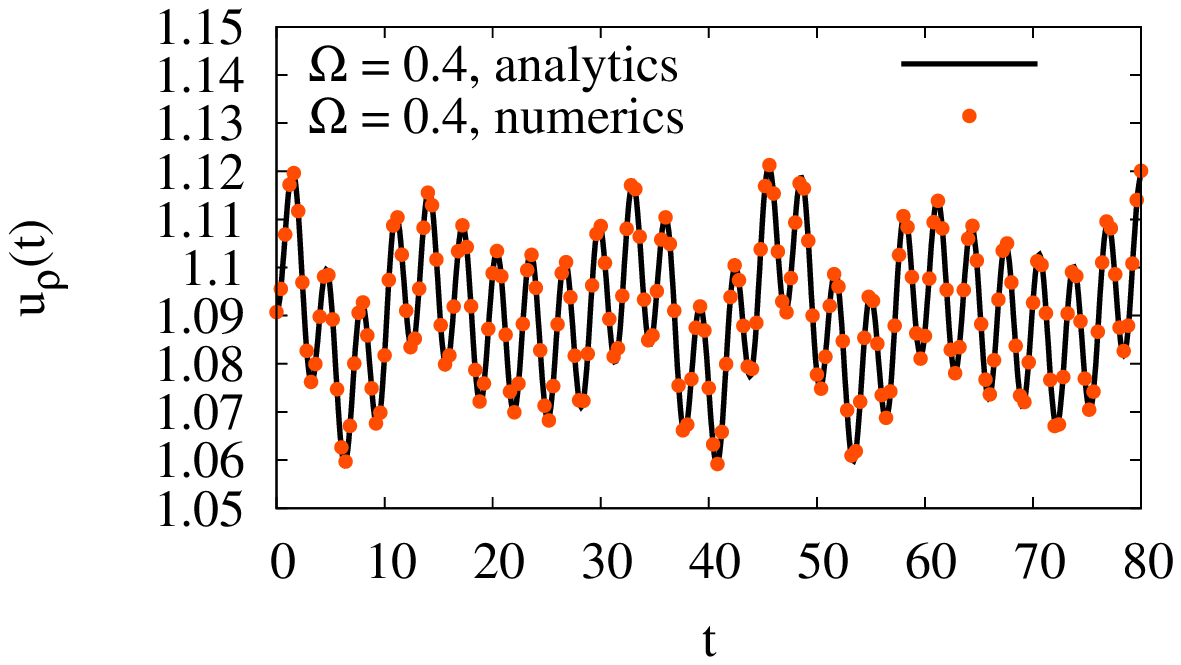}
\includegraphics[width=8cm]{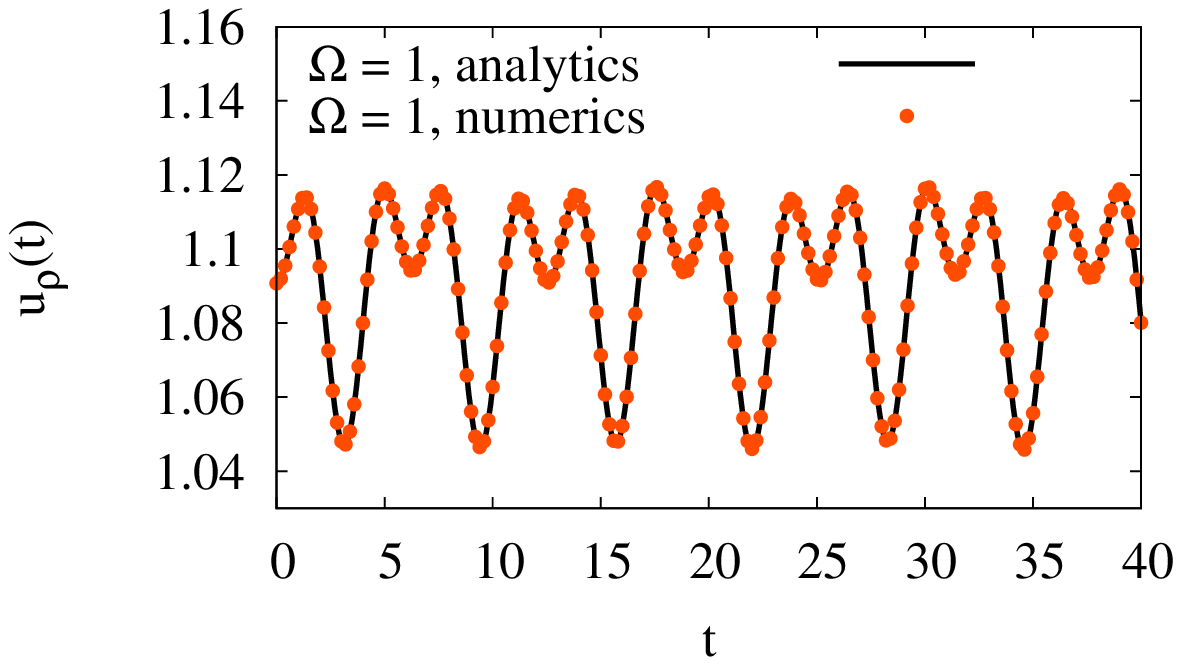}
\caption{(Color online) Condensate dynamics within the Gaussian approximation for $p=1$, $q=0.2$, $\lambda=0.3$ and two different driving frequencies $\Omega=0.4$ (left plot) and $\Omega=1$ (right plot). We plot exact numerical solution of  Eqs.~(\ref{eq:var1}) and (\ref{eq:var2}) together with the analytical second-order perturbative result, as explained in Section IV.A.
}
\label{fig:cynrt}
\end{center}
\end{figure*}

\begin{figure}[!hbt]
\begin{center}
\includegraphics[width=8cm]{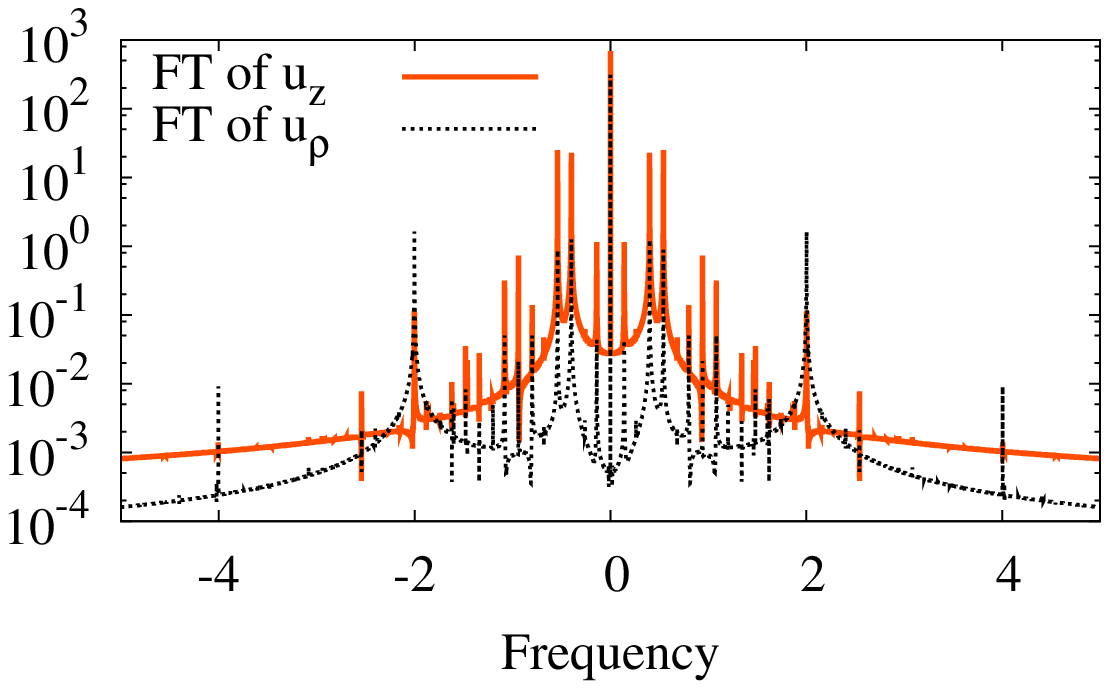}
\includegraphics[width=8cm]{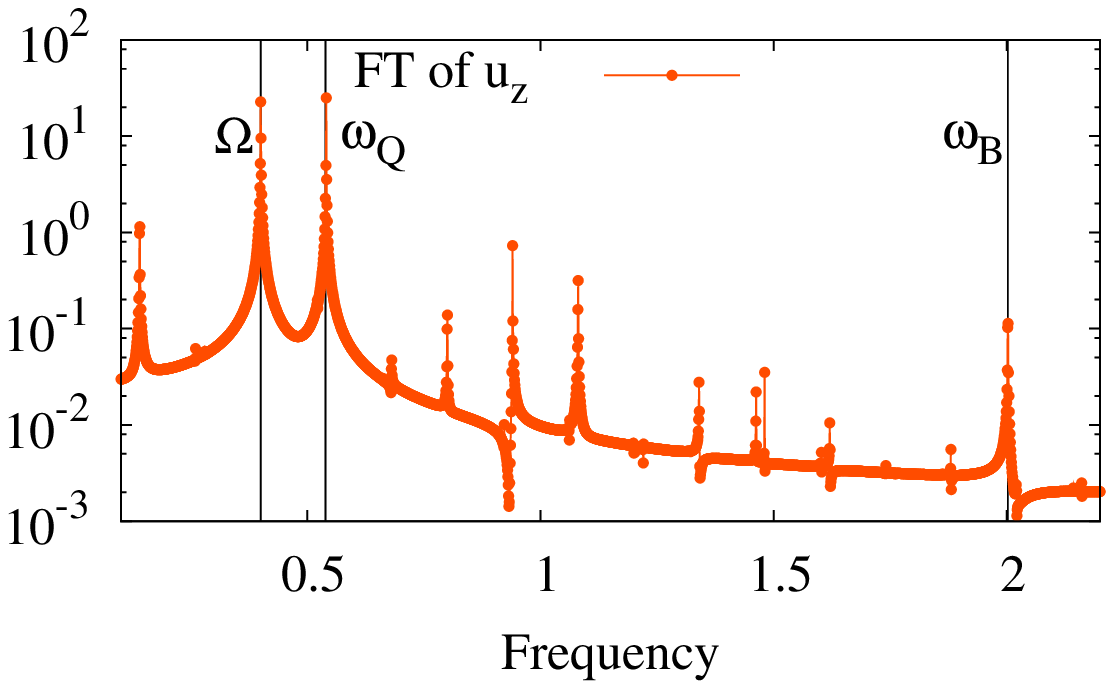}
\caption{(Color online) Fourier transformed $u_{\rho}(t)$ and $u_z(t)$ for $p=1$, $q=0.2$, $\lambda=0.3$, and $\Omega=0.4$. Upper plot gives complete spectrum, while on the lower plot we show part of the spectrum together with positions of prominent peaks.}
\label{fig:cynft}
\end{center}
\end{figure}

To obtain experimentally more relevant results, we now study an axially symmetric BEC.
An illustration of the condensate dynamics is shown in Fig.~\ref{fig:cynrt} for  $p=1$, $q=0.2$, $\lambda=0.3$. We plot numerical solutions of Eqs.~(\ref{eq:var1}) and (\ref{eq:var2}) obtained for the equilibrium initial conditions $u_{\rho}(0)=u_{\rho0}$, $\dot{u}_{\rho}(0)=0$, $u_{z}(0)=u_{z0} $, and $\dot{u}_{z}(0)=0$. For the specified parameters, the equilibrium widths are found to be  $u_{\rho0}=1.09073$, $u_{z0}=2.40754$ and from the linear stability analysis we find both the quadrupole mode frequency $\omega_{Q0}=0.538735$ and the breathing mode frequency $\omega_{B0}=2.00238$. For a driving frequency $\Omega$ close to $\omega_{Q0} $, we observe large amplitude oscillations in the axial direction. An example of excitation spectra is shown in Fig.~\ref{fig:cynft}. Here, we have the three basic modes $\omega_Q$, $\omega_B$, $\Omega$, and many higher-order harmonics.

\subsection{Poincar\'e-Lindstedt method}
In order to extract information on the frequencies of the collective modes beyond the linear stability analysis, we apply the perturbative expansion in the modulation amplitude $q$:
\begin{eqnarray}
\hspace{-5mm} u_{\rho}(t) &=& u_{\rho0} + q\, u_{\rho1}(t) + q^2\, u_{\rho2}(t) + q^3\, u_{\rho3}(t)+\ldots,\\
\hspace{-5mm}u_{z}(t) &=& u_{z0} + q\, u_{z1}(t) + q^2\, u_{z2}(t) + q^3\, u_{z3}(t)+\ldots,
\label{eq:cynexpansions} 
\end{eqnarray}
By inserting these expansions in Eqs.~(\ref{eq:var1}) and (\ref{eq:var2})  and by performing additional expansions in $q$, we obtain a system of linear differential equations of the general form:
\begin{eqnarray}
\ddot{u}_{\rho n}(t)+m_{11} u_{\rho n}(t)+m_{12} u_{z n}(t)+f_{\rho n}(t)&=&0,\label{eq:system1}\\
m_{21} u_{\rho n}(t)+\ddot{u}_{z n}(t)+m_{22} u_{z n}(t)+f_{z n}(t)&=&0,
\label{eq:system2}
\end{eqnarray}
where $n=1,2,3,\ldots$, $m_{11}=4$, $m_{12}=p/u_{\rho 0}^3u_{z0}^2 $, $m_{21}= 2\,p/u_{\rho 0}^3 u_{z0}^2$, and $m_{22}=\lambda^2+3/u_{z0}^4+2\,p/u_{\rho 0}^2 u_{z0}^3$. The functions $f_{\rho n}(t)$ and $f_{z n}(t)$ depend only on the solutions $u_{\rho i}(t)$ and $u_{z i}(t)$ of lower order, i.~e.~ $i<n$. For $n=1$ we have
$$
 f_{\rho1}(t)=-\frac{\cos \Omega t}{u_{\rho 0}^3 u_{z0}}, \quad f_{z1}(t)=-\frac{\cos \Omega t}{u_{\rho 0}^2 u_{z0}^2},
$$
and for $n=2$ we get correspondingly
\begin{eqnarray}
 & &\hspace{-10mm} f_{\rho2}(t)=\frac{3}{u_{\rho 0}^4 u_{z0}} \cos \Omega t\, u_{\rho 1}(t)-\frac{6 }{u_{\rho 0}^5}u_{\rho 1}(t)^2\nonumber\\
& &\hspace{3mm}-\frac{6 p }{u_{\rho 0}^5 u_{z0}}u_{\rho 1}(t)^2+\frac{1}{u_{\rho 0}^3 u_{z0}^2}\cos\Omega t\, u_{z1}(t)\nonumber\\
& &\hspace{3mm}-\frac{3 p}{u_{\rho 0}^4 u_{z0}^2} u_{\rho 1}(t) u_{z1}(t)-\frac{p}{u_{\rho 0}^3 u_{z0}^3}u_{z1}(t)^2,\nonumber\\
 & &\hspace{-10mm} f_{z2}(t)=\frac{2}{u_{\rho 0}^3 u_{z0}^2} \cos \Omega t\, u_{\rho 1}(t)-\frac{3p }{u_{\rho 0}^4 u_{z0}^2}u_{\rho 1}(t)^2\nonumber\\
& &\hspace{3mm}-\frac{3 p }{u_{\rho 0}^2 u_{z0}^4}u_{z1}(t)^2+\frac{2}{u_{\rho 0}^2 u_{z0}^3}\cos\Omega t\, u_{z1}(t)\nonumber\\
& &\hspace{3mm}-\frac{6}{u_{z0}^5} u_{z1}(t)^2-\frac{4p}{u_{\rho 0}^3 u_{z0}^3}u_{\rho 1}(t)u_{z1}(t).\nonumber
\end{eqnarray}
 The linear transformation 
\begin{eqnarray}
u_{\rho n}(t)&=&x_n(t)+y_n(t),\label{eq:lineartransform1}\\
u_{z n}(t)&=& c_1\, x_{n}(t)+c_2\, y_{n}(t),
\label{eq:lineartransform2}
\end{eqnarray}
with coefficients 
\begin{eqnarray}
c_1&=&\frac{m_{22}-m_{11}-\sqrt{(m_{22}-m_{11})^2+4m_{12}m_{21}}}{2 m_{12}},\nonumber\\ c_2&=&\frac{m_{22}-m_{11}+\sqrt{(m_{22}-m_{11})^2+4m_{12}m_{21}}}{2 m_{12}}\nonumber
\end{eqnarray} decouples the system at the $n$-th level and leads to:
\begin{eqnarray}
\ddot{x}_{n}(t)+\omega_{Q0}^2 x_n(t)+\frac{c_2 \, f_{\rho n}(t)-f_{z n}(t)}{c_2-c_1}&=&0,\label{eq:system21}\\
\ddot{y}_{n}(t)+\omega_{B0}^2 y_n(t)+\frac{c_1 \, f_{\rho n}(t)-f_{z n}(t)}{c_1-c_2}&=&0.
\label{eq:system22}
\end{eqnarray}
Now it is clear how to proceed: we first solve Eqs.~(\ref{eq:system21}) and ~(\ref{eq:system22}) for $x_1(t)$ and $y_1(t) $ and then using Eqs.~(\ref{eq:lineartransform1}) and (\ref{eq:lineartransform2}) we obtain $u_{\rho 1}(t)$ and $u_{z 1}(t)$. In the next step we use these solutions and solve for $x_2(t)$ and $y_2(t)$ and so on. At each level we impose the initial conditions $u_{\rho n}(0)=0$, $\dot{u}_{\rho n}(0)=0$, $u_{z n}(0)=0$, and $\dot{u}_{z n}(0)=0$. At the first level of perturbation theory, equations for $x$ and $y$ are decoupled, i.e. $x_1(t)$ and $y_1(t)$ are normal modes: $x_1(t)$ describes quadrupole oscillations, while $y_1(t)$ describes breathing oscillations. However, at the second order of perturbation theory $y_1(t)$ enters the equation for $x_2(t)$ and also $x_1(t)$ appears in equation for $y_2(t)$, i.e. we have a nonlinear mode coupling. 

The explicit calculation is performed up to the second order by using the software package Mathematica \cite{Mathematica}. We obtain a good agreement of analytical results obtained in the second order of our perturbation theory and numerical results, as can be seen in Fig.~\ref{fig:cynrt} for a moderate value of a modulation amplitude $q$. The first secular terms appear at the level $n=3$. The expressions are cumbersome, but the relevant behavior is obtained from the terms:
\begin{equation}
\ddot{x}_3(t)+\omega_{Q0}^2 x_3(t)+C_Q \cos(\omega_{Q0} t)+\ldots=0, 
\end{equation}
that leads to
\begin{equation}
x_3(t)=-\frac{C_Q}{2 \omega_{Q0}} t\sin(\omega_{Q0} t)+\ldots
\end{equation}
The last term can be absorbed into the first-order solution
\begin{eqnarray}
u_{\rho}(t)&=&A_Q \cos(\omega_{Q0} t) -\frac{C_Q q^2}{2 \omega_{Q0}} t\sin(\omega_{Q0} t)+\ldots\nonumber\\
&\approx&A_Q \cos(t(\omega_{Q0}+\Delta \omega_{Q0})),
\end{eqnarray}
and can be interpreted as a frequency shift of the quadrupole mode to the order $q^2$:
\begin{equation}
\omega_{Q} = \omega_{Q0}+\Delta \omega_{Q0}= \omega_{Q0}+q^2\frac{C_Q }{2 \omega_{Q0} A_Q}+\ldots 
\end{equation}
The coefficients $A_Q$ and $C_Q$ are calculated using the Mathematica code avaliable at our site \cite{scl}, but their explicit form is too long to be presented here. Along the same lines we calculate the frequency shift of the breathing mode.

\begin{figure}[!t]
\begin{center}
\includegraphics[width=8cm]{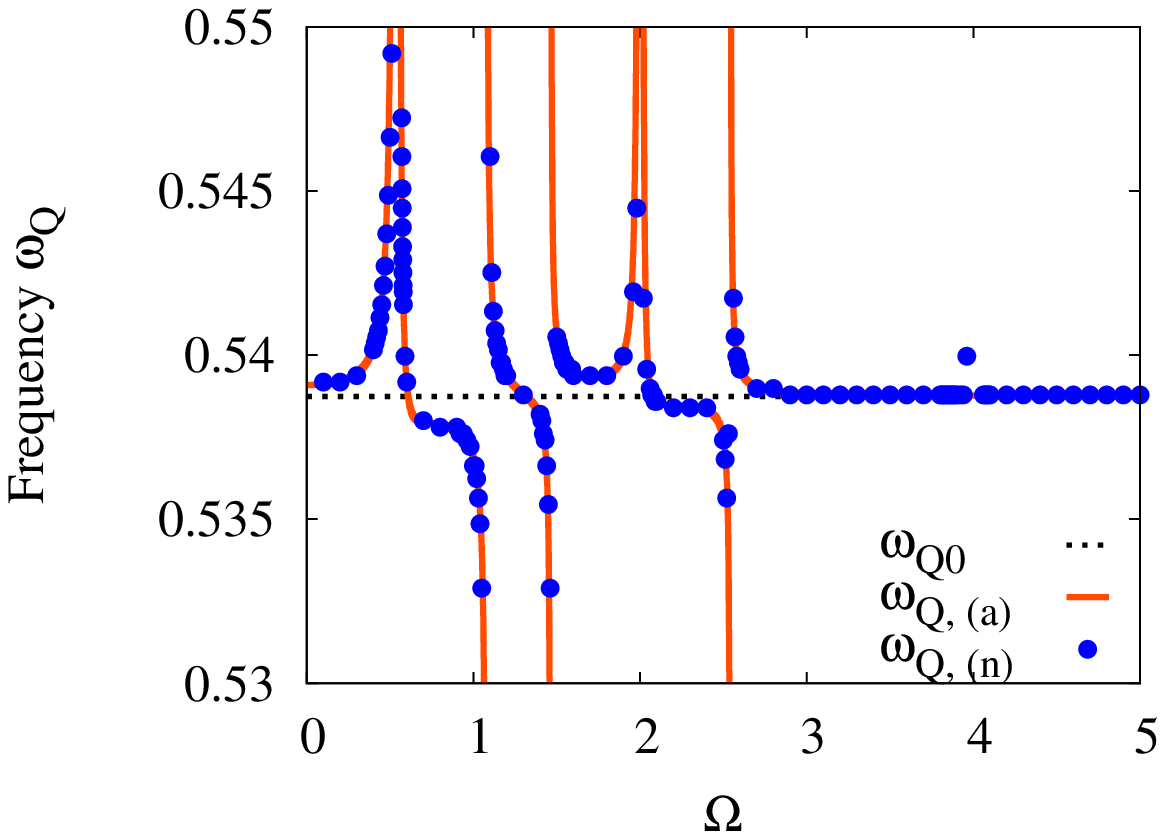}
\caption{(Color online) Frequency of the quadrupole mode $\omega_Q$ versus driving frequency
$\Omega$ for $p=1$, $q=0.2$, and $\lambda=0.3$. We plot linear response result $\omega_{Q0}$, second-order analytical result $\omega_{Q,\mathrm{(a)}}$ and numerical values $\omega_{Q,\mathrm{(n)}}$.}
\label{fig:cynqmshift}
\end{center}
\end{figure}
\begin{figure}[!ht]
\begin{center}
\includegraphics[width=8cm]{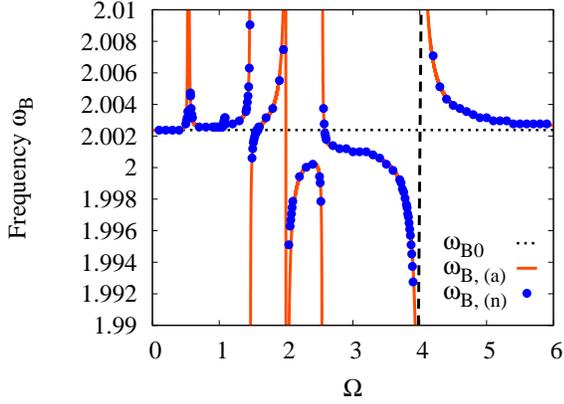}
\caption{(Color online) Frequency of the breathing mode $\omega_B$ versus driving frequency
$\Omega$ for $p=1$, $q=0.2$, and $\lambda=0.3$. We plot linear response result $\omega_{B0}$, second-order analytical result $\omega_{B,(a)}$ and numerical values $\omega_{B,(n)}$.}
\label{fig:cynbmshift}
\end{center}
\end{figure}

\subsection{Discussion}

The main results of our calculation are shown in Figs.~\ref{fig:cynqmshift} and \ref{fig:cynbmshift}.
In Fig.~\ref{fig:cynqmshift}  we plot the analytically obtained frequency of the quadrupole mode versus the driving $\Omega$, using the second order perturbation theory together with the corresponding numerical result based on the Fourier analysis of solutions of Eqs.~(\ref{eq:var1}) and (\ref{eq:var2}). 
 An analogous plot for the frequency of the breathing mode is given in Fig.~\ref{fig:cynbmshift}. Our analytical perturbative result for the shifted quadrupole mode frequency contains poles at $\omega_{Q0}$, $2\omega_{Q0}$, $\omega_{B0}-\omega_{Q0}$, $\omega_{Q0}+\omega_{B0} $ and $\omega_{B0}$. Similarly, for the shifted frequency of the breathing mode poles in the perturbative solution are found at $\omega_{Q0} $, $\omega_{B0} $, $2 \omega_{B0}$, $\omega_{B0}-\omega_{Q0}$ and $\omega_{Q0}+\omega_{B0}$. In both figures we see excellent agreement of the perturbatively obtained results with the exact numerics.

\begin{figure}[!b]
 \begin{center}
\includegraphics[width=8cm]{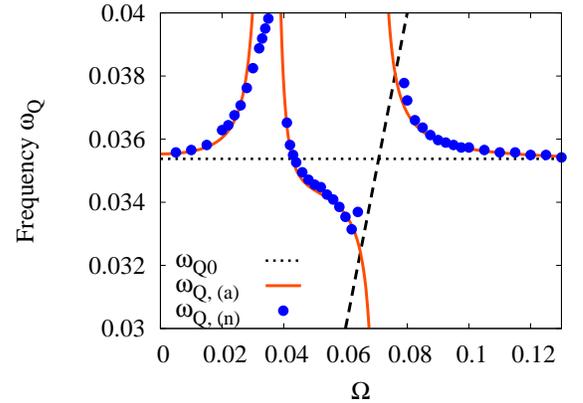}
\caption{(Color online) Frequency of the quadrupole mode $\omega_Q$ versus driving frequency
$\Omega$ for the experimental parameters from Eq.~(\ref{eq:exppar}). We plot linear response result $\omega_{Q0}$, second-order analytical result $\omega_{Q,(a)}$, and numerical values $\omega_{Q,(n)}$.}
\label{fig:exp}
\end{center}
\end{figure}

In the experiment \cite{parres}, excitations of a highly elongated and strongly repulsive BEC
were considered with the system parameters in Eq.~(\ref{eq:exppar}). For that case, according to Eq.~(\ref{eq:expfrequencies}) we get $\omega_{Q0}\ll\omega_{B0}$, and the driving frequency was chosen in the range $(0,3 \omega_{Q0})$. 
Good agreement of real-time dynamics obtained from the variational approximation with the exact solution of the time-dependent GP simulation occurs even for long propagation times according to Fig.~\ref{fig:gp}, which  implies a good accuracy of the Gaussian approximation for calculating the frequencies of the excited modes.
From the real-time dynamics plotted in Fig.~\ref{fig:gp} we observe the excitation of the slow quadrupole mode as an out-of phase oscillation in the axial and in the radial direction. In addition, in the radial direction we observe fast breathing mode oscillations. This is typical for highly elongated condensates \cite{salasnich} and our analysis for the experimental parameters shows a strong excitation of the quadrupole mode, but also a significant excitation of breathing mode in the radial direction. Due to the large modulation amplitude $q$, many higher order harmonics are excited, and, most importantly, we find frequency shifts of the quadrupole mode of about 10\% in Fig \ref{fig:exp}. From the same figure we notice that, due to the chosen frequency range for $\Omega$, only resonances located at $\omega_Q $ and $2\omega_Q$ are observed. The presence of nonlinear effects is already mentioned in Ref.~\cite{parres}. However we suggest frequency shifts calculated here to be taken into account for extracting the  resonance curves from the underlying experimental data. 

To achieve more clear-cut experimental observation of the nonlinearity-induced frequency shifts calculated in this paper, we suggest a different trap geometry from the one used in Ref.~\cite{parres}. Measurements of stable BEC modes can be performed for about $1\,\mathrm{s}$ and in order to extract precise values of the excited frequencies in the Fourier analysis, several oscillation periods should be captured within this time interval. A higher frequency of the quadrupole mode, that can be realized by using a larger trap aspect ratio $\lambda$, in combination with a higher modulation frequency would fulfill this condition. According to the results presented in Ref.~\cite{parres}, resonant driving may lead to condensate fragmentation.
However, our numerical results indicate frequency shifts of $10\,\%$ even outside the resonant regions according to Figs.~\ref{fig:sphresult} and \ref{fig:exp}, and this is where experimental measurements should be performed. Although an increase in $\lambda$ leads to a more pronounced nonlinear mixing of quadrupole and breathing mode and may complicate condensate dynamics further, it may be possible to perform a Fourier analysis of experimental data, analogous to Ref.~\cite{fortagh}, and to compare it with the excitation spectra presented here. To achieve a complete matching of experimental data and our calculations, it may turn out that higher-order corrections to Eq.~(\ref{eq:tdsl}), that arise due to nonlinear dependence of scattering length on the external magnetic field, have to be taken into account.

\section{Conclusions}

Motivated by recent experimental results, we have studied nonlinear BEC dynamics
induced by a harmonically modulated interaction at zero temperature. We have used a combination of an analytic perturbative approach, numerical analysis based
on Gaussian approximation, and numerical simulations of a full time-dependent Gross-Pitaevskii equation.
We have presented numerically calculated relevant excitation spectra and found prominent nonlinear
features: mode coupling, higher harmonics generation, and
significant shifts in the frequencies of collective modes. In addition, we have provided an analytical perturbative framework that captures most of the observed phenomena. The main results are analytic formulae describing the dependence of collective mode frequencies on the modulation amplitude and on the external driving frequency for different trap geometries.
To extend the applicability of our analytical approach, a perturbative expansion to higher order has to be performed, or an appropriate resummation of the perturbative series could be applied.  

The presented results could contribute to future experimental designs that may include
mixtures of cold gases and their dynamical response to harmonically modulated
interactions, such as pattern formation induced by the modulation of different time dependence  of the scattering length.
In addition, our results could contribute to resolving beyond-mean-field effects in the collective mode frequencies, as proposed in Refs.~\cite{beyondmeanfield1, beyondmeanfield2}, and for dipolar BEC in \cite{Lima&Pelster}. Nonlinearity-induced shifts of collective modes have to be properly taken into account to clearly delineate them from beyond-mean-field effects. 

\begin{acknowledgments}
We thank Vanderlei Bagnato for many discussions concering the experimental results presented in Ref.~\cite{parres}.
This work was supported in part by the Ministry of Education and Science of the Republic of Serbia
 under projects No.~ON171017 and  NAD-BEC, by DAAD - German Academic and Exchange Service under project NAD-BEC, and by the European Commission under EU FP7 projects PRACE-1IP, HP-SEE and EGI-InSPIRE.
\end{acknowledgments}

\providecommand{\noopsort}[1]{}\providecommand{\singleletter}[1]{#1}%

\end{document}